\documentclass[pre,12pt,english]{revtex4-2}
\usepackage[T1]{fontenc}
\usepackage[latin9]{inputenc}
\usepackage[fleqn]{amsmath}
\usepackage{mathptmx}
\usepackage{amssymb}
\usepackage{mathtools}
\usepackage{graphicx}
\usepackage{epstopdf}
\usepackage{esint}
\usepackage{color}
\usepackage{ulem}

\usepackage{lineno,hyperref}
\modulolinenumbers[5]
\usepackage{babel}

\begin{document}

\begin{abstract}
  We propose a model for  the motion of a single active particle in a heterogeneous environment where the heterogeneity may arise due to the crowding, conformational fluctuations and/or slow rearrangement of  the surroundings. Describing the active particle in terms of the Ornstein-Uhlenbeck process (OUP) and incorporating the heterogeneity in the thermal bath using the two separate models, namely ``diffusing diffusivity" and ``switching diffusion", we explore the essential dynamical properties of the particle  for  its one-dimensional motion. In addition, we show how the dynamical behavior is controlled by dynamical variables associated with the active noise such as  strength and  persistence time. Our model is relevant in the context of single particle dynamics in crowded environment, driven by activity.  
\end{abstract}

\title{Motion of  an active particle with dynamical disorder}	
\author{Koushik Goswami* and Rajarshi Chakrabarti**}
\address{Department of  Chemistry, Indian Institute of Technology Bombay, India\\  *E-mail: goswamikoushik10@gmail.com\\ **E-mail: rajarshi@chem.iitb.ac.in}

\address{\textbf{Published in Soft Matter, 2022, 10.1039/D1SM01816G}}

\maketitle
\section{Introduction}
\noindent Recent  advances in the single-particle tracking technique have promoted the studies on the motion of single particles which are of great importance as they provide  deep insights about the nonequilibrium physics, in addition to their immense potential applicability in the field of biophysics \cite{thompson2010three,shen2017single}. In biological systems, from cargo transport by a motor protein to the motion of a microorganism such as bacteria, a particle moves on its own by harnessing energy from an active process like ATP hydrolysis. Such system is termed as active  and it operates  out of equilibrium \cite{ramaswamy2010mechanics,fodor2018statistical}.  
One of the most common ways of modelling an active particle is by describing its propulsion velocity as the Ornstein-Uhlenbeck process (OUP), and it is termed as the  ``active Ornstein-Uhlenbeck particle" (AOUP) model \cite{maggi2014generalized,PhysRevE.100.022601,martin2021statistical,D0SM01200A}.   In this model, the  velocities at different times  are  exponentially correlated \cite{PhysRevE.100.022601,um2019langevin}. Therefore, the motion has  persistence. By construction, in the AOUP framework, the particle has a Gaussian distribution in space, and it ensures enhanced diffusion in the long-time limit. It is worth mentioning here that the OUP model can also be adapted to describe the dynamics of passive particles immersed in a bath containing active particles, $e.g.,$ see Refs.  \cite{chaki2018entropy,goswami2019heat,goswami2019work,chaki2019effects,goswami2021work}.  The above description is  well suited for the diffusion in an environment where the diffusivity of the particle  remains unchanged over the time, or in other words, the particle moves in a homogeneous environment.

\noindent Over the past decade, experiments on the dynamics of  passive tracers in a crowded environment as well as computer simulations have been performed by several research groups \cite{wang2009anomalous,guan2014even,kwon2014dynamics,kumar2019transport,chakraborty2019nanoparticle,skaug2013intermittent,PhysRevResearch.2.022020}. The examples include diffusion of lipid vesicles in a solution of entangled filaments \cite{guan2014even},  tracer diffusion in an environment consisting of large particles or polymer melts or suspension of polymer chains \cite{kwon2014dynamics,xue2016probing,samanta2016tracer}, mobile biomolecules on a surface \cite{skaug2013intermittent,chakraborty2019nanoparticle}, in-plane diffusion of drug molecules in between silica slabs \cite{D0CP03849K},  to name but a few. The most notable observation of these studies is the fact that  particle has non-Gaussian position distribution.  In many cases, the distribution has an exponential tail at intermediate times, which vanishes in the long-time limit, exhibiting a  Gaussian behavior, as predicted by the central limit theorem \cite{wang2012brownian,samanta2016tracer}.  Interestingly, some studies have shown that though the distribution is non-Gaussian, the mean square displacement (MSD)  all times grows linearly in time, or in other words, the dynamics is Fickian \cite{wang2009anomalous,guan2014even,kwon2014dynamics,PhysRevResearch.2.022020}.  Such observations are not consistent with the description of  the normal diffusion, and thus the processes associated with similar features are, in general, termed as ``anomalous, yet Brownian diffusion" (AYBD) \cite{PhysRevLett.113.098302,PhysRevLett.124.060603}.  A plausible explanation for AYBD comes from a  theoretical  formalism where the tagged particle possesses multiple diffusion coefficients, $i.e.,$ the diffusivity is a stochastic variable  and  it follows some distribution \cite{PhysRevLett.113.098302}.  The idea of introducing fluctuations into intensive variables such as energy dissipation can be traced back to the work on atmospheric turbulence  by Obukhov and Kolmogorov  \cite{https://doi.org/10.1029/JZ067i008p03011,oboukhov1962some,kolmogorov1962refinement,naert1994velocity}.  Later similar idea was invoked by Beck and Cohen to show how an effective statistical description can be obtained after averaging over the  fluctuations of intensive variables such as temperature (or thermal diffusivity), and they  called it  ``superstatistics" \cite{beck2003superstatistics}.   However, it was Robert Zwanzig who, in a different context, introduced essentially the same concept in a more general way while  describing the  rate process with a rate constant that fluctuates in time due to fluctuations of the barrier height  \cite{zwanzig1990rate}. It can take discrete values or it can be a continuous function. Zwanzig called this    ``dynamical disorder"  \cite{doi:10.1063/1.2200695,samanta2016tracer,tyagi2017non}. In the context of diffusing particles,  a  similar formalism named ``diffusing  diffusivity" (DD) model was put forward by  Chubynsky and Slater \cite{PhysRevLett.113.098302}, and it was further extended by Sebastian's group  and Metzler's group independently \cite{doi:10.1021/acs.jpcb.6b01527,jain2016diffusing,PhysRevE.95.032135,chechkin2017brownian,metzler2017gaussianity,sposini2018random}. The model of Chubynsky and Slater is a case of static (or quenched) disorder  as the diffusivity does not fluctuate with time as it has a stationary distribution.  But  for the models developed by Sebastian  and Metzler,  the diffusivity of a  particle in a crowded, heterogeneous surroundings is a time-dependent random variable which is considered as the square of a random process such as OU process, since the diffusivity can take only positive values. But when the motion  is sampled over all trajectories, it  results diffusive motion with the ensemble-averaged diffusivity.  In other words, the mean square displacement is proportional to the time elapsed, $i.e.$, $\langle x^2(t)\rangle=K t$, and the coefficient of  proportionality ($K$) is fully time-independent quantity which can be expressed in terms of the equilibrium diffusivity.  This model is applied to describe the  motion in  a heterogeneous environment rearranging with a  timescale comparable to the one corresponding to the diffusing particle.  It should  be noted here that in the quenched-disordered media the DD approach can also be applied but only in the limit of rapidly changing disorder  \cite{PhysRevE.97.042122,PhysRevE.100.042136,Postnikov_2020}.   For different random-diffusivity models, readers are referred to Refs. \cite{lanoiselee2018diffusion,sposini2020universal,slkezak2021diffusion}. See also Ref. \cite{PhysRevLett.98.250601} for the derivation of AYBD using the large deviations approach.  There are situations where the particle can have few  conformations specified by their diffusivities, or there are spatial heterogeneity or specific kinds of interaction between the particle and its surroundings which leads to a motion associated with discrete diffusivities \cite{PhysRevLett.81.4915,leith2012sequence,parry2014bacterial,cuculis2015direct}. For instance, a freely diffusing particle intermittently trapped due to presence of random binding zones or traps exhibits a non-Gaussian spatial distribution \cite{PhysRevE.98.040101}. Another  example includes conformational fluctuations of  a protein called transcription factor which switches between two conformations while searching for a promoter site to initiate the transcription; in one conformation it moves fast on the DNA track, but in another conformation it undergoes a slow motion due to strong interaction with the track \cite{leith2012sequence}.  The above dynamics  is usually described  by   the two-state model or the switching diffusion  \cite{PhysRevE.94.012109,tyagi2017non}. Apart from the Brownian dynamics, applications of the switching model in describing several biochemical processes such as cellular signalling, chemotaxis, synaptic dynamics,  growth of cell population,  pattern formation are noteworthy in relation to the present topic \cite{bressloff2017stochastic,PhysRevLett.108.158101,godec2017first,paul2018reaction,PhysRevE.104.L012601}.  In addition to the AYB diffusion of Brownian particles, stochasticity in diffusivity   naturally arises in the dynamics of macromolecules  such as  conformational fluctuations of  proteins \cite{PhysRevLett.126.128101,PhysRevE.104.L062501} and the motion of  the center of mass of (de)polymerising or shape-shifting molecules \cite{10.3389fphy.2019.00124,PhysRevE.104.L062501}.  The non-Gaussian behavior can also be observed for sub-diffusing particles moving in gels or viscoelastic media such as cytoplasmic environment due to  heterogeneity \cite{lampo2017cytoplasmic,cherstvy2019non,wang2020unexpected,PhysRevE.98.022122}. See the theoretical work of  Ref. \cite{PhysRevX.11.031002} where unexpected non-Gaussianity occurs for Lennard-Jones mixtures as a result of intermittent hopping motion.

\noindent In all the examples discussed above on the  heterogeneous diffusion, the particle's motion is mainly influenced by the thermal fluctuations, or more specifically, no intrinsic energy source or no external stochastic driving is considered, and therefore, the dynamics is in equilibrium.  However, as mentioned earlier, the diffusive motion coupled to an active process is very common in biological systems, though its theoretical study in heterogeneous environment is largely  lacking. Inclusion of  heterogeneity  is important to study the activity-driven systems such as  self-propelled particles in the presence of obstacles \cite{PhysRevLett.110.238101}, active soft colloids with switching active interactions \cite{D1SM01507A},  run-and-tumble disks displaying avalanche dynamics \cite{reichhardt2018avalanche}, transport of active tracers in a polymer grafted channel \cite{D1SM01693H}, etc. For a detailed account on this topic, the reader is referred to Ref.  \cite{RevModPhys.88.045006}. Along these lines, there have been theoretical investigations of  the impact of heterogeneity on the transport properties of active particles \cite{thiffeault2021shake,PhysRevE.104.064615}. These studies have mostly considered the active Brownian particle (ABP) model to describe the dynamics, and the fluctuations are incorporated either in self-propulsion velocity or both in thermal and active forces. In another study  \cite{D1SM01507A},  Bley \textit{et. al.} have investigated the dynamical properties of a suspension of active colloids where each particle switches randomly between two states differing in their size as well as  mobility. It has been shown there  that this switching induces non-Gaussian behavior at  intermediate times.

\noindent Drawing motivations from these experimental and simulation based investigations,  here we propose an analytically solvable model to capture  the dynamics of a simple one-dimensional active particle with dynamical disorder.  We consider two cases, namely the switching diffusion where the particle switches between two conformations with two different thermal diffusivities, and the diffusing diffusivity model in which  the particle  takes instantaneous thermal diffusivity which changes randomly over time due to the rearrangement of the environment. In our model, the activity does not get affected by the heterogeneity, and it is modeled as Ornstein-Uhlenbeck process. These models along with the dynamics of the particle is discussed in Sec. \ref{dynamics}. In Sec. \ref{results} we showcase  the main results. A summary is given in Sec. \ref{conclusion}.  
\section{Dynamics \label{dynamics}}
\begin{figure}
\centering
\includegraphics[width=0.5\linewidth]{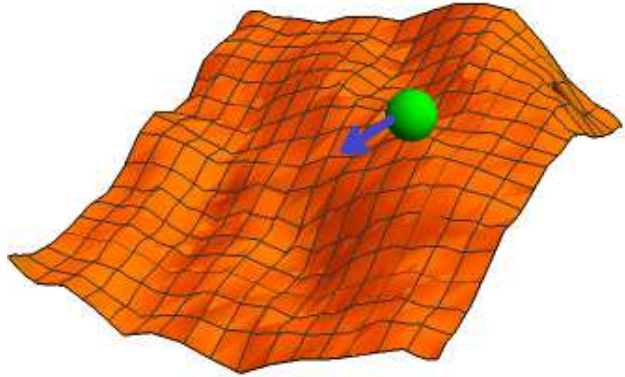}
\caption{A typical illustration of a diffusing active particle (green ball) in a disordered or heterogeneous medium. The blue arrow represents the instantaneous direction of the self-propulsion velocity ($v_A$) of the particle. }
   \label{model_pic}
\end{figure}

\noindent Here we consider an active particle moving  in a heterogeneous media as pictorially depicted in Fig. \ref{model_pic}. For simplicity, the motion is restricted to one dimension although it can  be in principle extended to higher dimensions. In the following section, more details about the dynamics is provided.

\subsection{Model of active noise \label{model-active}}
\noindent Apart from thermal kicks experienced by the particle, it is also subjected to an extra noise termed as the active noise, and as a result, it executes directed motion. Therefore, the dynamics of the particle  is governed by the following stochastic equation: 
 \begin{align}
\dot{x}(t)=\eta_T(t)+v_A(t),\label{langevin}
 \end{align}
 where $\eta_T(t)$ is the thermal noise modeled as a Gaussian white noise with zero mean. Thus, $\langle \eta_T(t) \rangle=0,$ and the autocorrelation function is  
 \begin{align}
\langle \eta_T(t) \eta_T(t') \rangle=2D_T(t)\delta(t-t').\label{white_corr}
 \end{align}
 Here it is assumed that either the environment is crowded and stochastically rearranging, that is to say, the particle explores the heterogeneous environment,  or the particle itself alters between many states, which results only the fluctuations in the thermal diffusivity $D_T.$ Following Zwanzig, it is referred here as the dynamically disorder system. To  account for the disorder, two different models are discussed in Sec. \ref{model-diff}.
 
\noindent On the other hand, $v_A(t)$ corresponds to the self-propulsion velocity of the particle, and it can be conceived as the active noise which  drives the particle out of equilibrium. As a  standard model, the noise  taken here is a Gaussian colored noise, which  follows the OUP of the form
\begin{align}
\dot{v}_A(t)=-\frac{1}{\tau_A}v_A(t)+\frac{1}{\tau_A}\eta_A(t),\label{color_corr0}
\end{align}
with $\eta_A(t)$ being the Gaussian white noise  having correlation $\langle \eta_A(t)\eta_A(t')\rangle=2 D_A \delta(t-t').$ So the auto-correlation function of $v_A(t)$ can be expressed as 
\begin{align}
\langle v_A(t)v_A(t') \rangle=\frac{D_A}{\tau_A} e^{-\frac{|t-t'|}{\tau_A}},\label{color_corr}
\end{align} 
where $D_A$ denotes the active diffusivity and $\tau_A$ is the persistence time. Other than the strength $D_A,$ the characteristic timescale $\tau_A$ characterizes the active noise, and  higher value of $\tau_A$ means a long-lived directed motion. In the limit $\tau_A \rightarrow 0,$ the noise becomes delta-correlated and behaves like the thermal noise \cite{chaki2020escape}. 

\subsection{Model of thermal diffusivity \label{model-diff}}
\noindent As discussed earlier, the thermal diffusivity of the particle fluctuates in time due to disorder. In the following sections, two models are considered to describe the time-dependent diffusivity $D_T(t).$ 
\subsubsection{Switching diffusion \label{sw_diff}}
\noindent  In the switching diffusion model,  the particle switches between two states with two different thermal diffusivities $D_1$ and $D_2.$ Let us call the states as state $1$ and state 2, respectively.  The transition happens from the state 2 to  the state 1  with a rate $r_{21},$ and vice versa with a rate $r_{12}.$ This is a simple version of the   ``Markov additive" model commonly used to capture the protein movement on a DNA track where it encounters different states of chromatin with variable affinities, and as a result, its motion accounts for the heterogeneous environment \cite{garcia2021nar}.  Here we  assume that the particle has two discrete states and it starts its journey either from the state 1 or state 2 with their respective steady-state probabilities which are given by
\begin{align}
 p_{1,s}=\frac{r_{21}}{r_{12}+r_{21}},\quad p_{2,s}=\frac{r_{12}}{r_{12}+r_{21}},\label{weight_st}
\end{align}  where $p_{i,s}$ being the steady-state probability at state $i$, $i=1,\,2.$  For such a case, the characteristic functional can be  expressed as  \cite{PhysRevE.102.042103}
\begin{align}
\langle e^{-p^2\int_{0}^{t}dt'\,D(t')}\rangle_{D(t')}=\frac{1}{2}\left[1-\frac{\phi_3}{\phi_2}\right]\,e^{-(\phi_1+\phi_2)t}+\frac{1}{2}\left[1+\frac{\phi_3}{\phi_2}\right]\,e^{-(\phi_1-\phi_2)t},\label{characterisctic-dicho}
\end{align}
where $\phi_1=\frac{p^2}{2}(D_1+D_2)+\frac12(r_{12}+r_{21}),\,\phi_2 =\frac{1}{2}\sqrt{\left[(D_2-D_1)p^2 +(r_{21}-r_{12})\right]^2+4r_{12}r_{21}},$ and $\phi_3 =\frac12 (r_{12}+r_{21})-\frac{p^2}{2} \frac{r_{12}-r_{21}}{r_{12}+r_{21}}(D_2-D_1).$ The equilibrium diffusivity is given by  
$D_{eq}=\langle D(t)\rangle=p_{2,s}D_2+p_{1,s}D_1.$ Here, $D_2\geq D_1 \geq 0.$

\subsubsection{Diffusing diffusivity model \label{DD_model}}
\noindent In the diffusing diffusivity model,  $D_T(t)$ is considered as the position of $n-$dimensional harmonic oscillator, $i.e.$, $D_T(t)=\sum_{i=1}^{n}\xi_i^2(t),$ where the evolution of  $\xi_i(t)$ is governed by  the Ornstein-Uhlenbeck process (OUP), 
 $\dot{\xi}_i=-\omega\,\xi_i(t)+\eta_i(t).$ Here   $\eta_i(t)$ is the white Gaussian noise with correlation $\langle \eta_i(t) \eta_j(t')\rangle=D_{eq}\omega\,\delta_{ij}\,\delta(t-t'),$  and $D_{eq}$ is the effective or equilibrium diffusivity of the particle in the medium.
 Here, we consider a two-dimensional model, $i.e.$, $D_T(t)=\sum_{i=1}^{2} \xi_i^2(t).$ The characteristic functional for such model can be found easily by the path integral technique, and it reads \cite{doi:10.1021/acs.jpcb.6b01527}
\begin{align}
\langle e^{-p^2\int_{0}^{t} dt_1\,D(t_1)}\rangle_{D(t_1)}=\frac{4 \omega \beta\,e^{-(\beta-\omega) t}}{(\beta+\omega)^2-(\beta-\omega)^2 \,e^{-2\beta t}},\label{dangle}
\end{align} 
with $\beta=\sqrt{\omega^2+2 \omega D_{eq} p^2}.$ The relaxation timescale for the environment rearrangement is roughly related to the inverse of  $\omega,$ which implies that for higher values of $\omega,$ the environment relaxes faster  and as a result, the particle visits all possible  configurations of the environment comparatively at a shorter timescale. 
\section{Results \label{results}}

\noindent Here our aim is to find the dynamical properties of the particle. To do so,  we  first outline the technique to obtain the probability distribution function (PDF) in space. Using Eq. (\ref{langevin}), one can write 
\begin{align}
    x(t)=x_0+\int_{0}^{t}dt_1\,\eta_T(t_1)+\int_{0}^{t}dt_1\,v_A(t_1),\label{xt}
\end{align}
with $x_0$ being the initial position of the particle. 
Now the probability distribution function (PDF) of finding the particle at position $x$ at time $t$  considering $x(0)=x_0,$ can be expressed as  
 \begin{align}
P(x,t|x_0,0)&=\langle \delta(x-x(t))\rangle,
 \end{align}
where $\langle \cdots\rangle$ represents the ensemble average over all trajectories of $x(t).$ Writing   the definition of delta functional explicitly in the above and with the aid of Eq. (\ref{xt}), the PDF can be rewritten as 
 \begin{align}
&P(x,t|x_0,0)\nonumber\\
&=\frac{1}{2\pi}\int_{-\infty}^{\infty}dp\, e^{-i p x}\langle e^{i p x(t)}\rangle\nonumber\\
&=\frac{1}{2\pi}\int_{-\infty}^{\infty} dp\, e^{-i p (x-x_0)}\langle e^{i p \int_{0}^{t}dt'\,\eta_T(t')} \rangle_{\eta_T(t')}\,\langle e^{i p \int_{0}^{t}dt'\,v_A(t')} \rangle_{v_A(t')}.\label{P_AOUP1}
 \end{align}
 Now the averages are taken over the  noises. As the evolution of the two noises are decoupled to each other, the ensemble averages now can be computed separately as shown in the above equation.  
It is useful to note here that the fourth-order autocorrelation function of any Gaussian noise $\sigma_G$ can be expressed as \cite{risken1996fokker}
\begin{align}
 &\langle\sigma_G(t_1)\sigma_G(t_2)\sigma_G(t_3)\sigma_G(t_4)\rangle\nonumber\\
 &=\langle\sigma_G(t_1)\sigma_G(t_2)\rangle\langle\sigma_G(t_3)\sigma_G(t_4)\rangle + \langle\sigma_G(t_1)\sigma_G(t_3)\rangle\langle\sigma_G(t_2)\sigma_G(t_4)\rangle \nonumber\\
 &\;+\langle\sigma_G(t_1)\sigma_G(t_4)\rangle\langle\sigma_G(t_3)\sigma_G(t_2)\rangle, \label{fourthmonent} 
\end{align}
 and the characteristic functional of the noise $\sigma_G$ can be evaluated using the relation \cite{goswami2021nonequilibrium}
\begin{align}
 \langle e^{i  \int_{0}^{t}dt'\,p(t')\sigma_G(t')} \rangle_{\sigma_G(t')}= e^{-\frac12\int_{0}^{t}dt_1\int_{0}^{t}dt_2 p(t_1)\langle\sigma_G(t_1)\sigma_G(t_2)\rangle p(t_2)}.
\end{align}
By the virtue of the above relation and  Eq. (\ref{color_corr}), the average over the  active noise can be easily obtained, and it reads 
\begin{align}
 \langle e^{i p \int_{0}^{t}dt'\,v_A(t')} \rangle_{v_A(t')}= e^{-\frac{p^2}{2}\int_{0}^{t}dt_1\int_{0}^{t}dt_2\langle v_A(t_1)v_A(t_2)\rangle}=  e^{- D_A p^2 \left(t+\tau_A \left(e^{-\frac{t}{\tau _A}}-1\right)\right)}.\label{ch_active}
\end{align}
Similarly for the thermal noise, the average can be calculated using Eq. (\ref{white_corr}) as 
\begin{align}
 \langle\langle e^{i p \int_{0}^{t}dt'\,\eta_T(t')}\rangle \rangle_{\eta_T, D_T}=\langle e^{-p^2 \int_{0}^{t}dt'\,D_T(t')}\rangle_{D_T(t')}. 
\end{align}
Since the the diffusivity is also a stochastic variable, one also needs to do the averaging over all realizations of $D_T(t),$ as  shown in the above equation. Now we can take  different models of $D_T(t)$ as discussed in Sec. \ref{model-diff}  to do the further analysis. 

\begin{figure}
\centering
\includegraphics[width=0.5\linewidth]{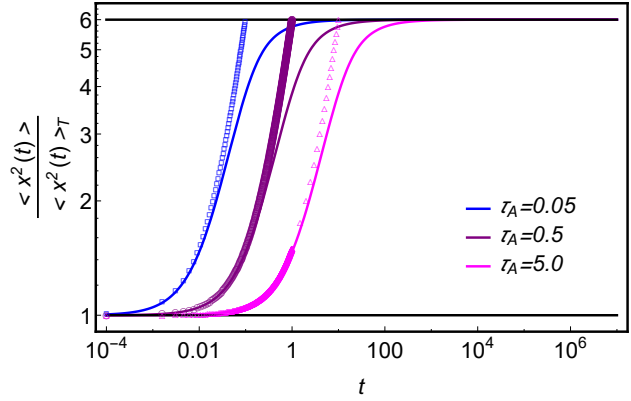}
\caption{Log-Log plot of the mean square displacement (MSD) [Eq. (\ref{msd})] scaled by the MSD in a purely thermal bath as a function of time $t$ for different values of $\tau_A.$ The curves with symbols are the approximate results for the ballistic regime in the limit $t\ll \tau_A$ [see Eq. (\ref{approx_msd})], and two black lines correspond to the very short-time (lower line) and long-time (upper line) diffusive limits. The other parameters used in the plots are $D_{eq}=5.0,\,D_A=25,\,x_0=0.$ }
\label{plot_msd}
\end{figure}

\noindent To get a preliminary idea about the dynamics, we find the mean square displacement (MSD). For both the models of $D_T(t),$ the MSD can be expressed using Eq. (\ref{xt}), and it yields (see Eq. (\ref{msd11}))
\begin{align}
&\langle (x(t)-x_0)^2\rangle\nonumber\\&=\int_{0}^{t}\int_{0}^{t}dt_1 dt_2\,\langle\eta_T(t_1)\eta_T(t_2)\rangle+\int_{0}^{t}\int_{0}^{t}dt_1 dt_2\,\langle v_A(t_1)v_A(t_2)\rangle \nonumber\\
&=2\,D_{eq}\, t+2\,D_A \left(t+\tau _A \left(e^{-\frac{t}{\tau _A}}-1\right)\right)\label{msd}.
\end{align}
 Notice that the first term of the right-hand side (RHS) is a linear function of time and is solely coming from the thermal contribution, which we denote here as $\langle (x(t)-x_0)^2\rangle_T.$   Now we can distinguish different time limits for which the MSD can be approximated as follows:
\begin{align}
 \langle (x(t)-x_0)^2\rangle\approx \begin{cases}
  \begin{cases}
   2 D_{eq}\, t  & \text{if $t< 2 \frac{D_{eq}}{D_A}\tau_A$} \\
 D_A\frac{t^2}{\tau_A}  & \text{if $t> 2 \frac{D_{eq}}{D_A}\tau_A$}
 \end{cases}
 & \text{ if $t\ll \tau_A$} \\
 \begin{cases}
 2 (D_{eq}+D_A) t 
 \end{cases}
 & \text{if $t\gg \tau_A$}
 \end{cases} \label{approx_msd}.
\end{align}
In the short-time limit, $i.e.,$ for $t\ll \tau_A,$ the motion approaches a ballistic behavior when $t> 2 \frac{D_{eq}}{D_A}\tau_A,$ though at $t<2 \frac{D_{eq}}{D_A}\tau_A,$ the dynamics is Fickian  with the thermal diffusivity $D_{eq}.$ On the other hand, the MSD for the long time limit $t\gg \tau_A$ is linear in time, clearly suggesting a diffusive regime with an elevated diffusivity. The results are graphically displayed in Fig. \ref{plot_msd} and are equivalent to the one observed for the case in a homogeneous environment \cite{goswami2019diffusion}.

\noindent To know more about the dynamics, particularly the tail behavior, we compute the non-Gaussian parameter (NGP) defined as $\gamma_G=\frac{ \langle x^4(t)\rangle}{3\langle x^2(t)\rangle^2}-1,$ considering $x_0=0,$ without any loss of information. 

\begin{figure}
 \centering
\includegraphics[width=0.45\linewidth]{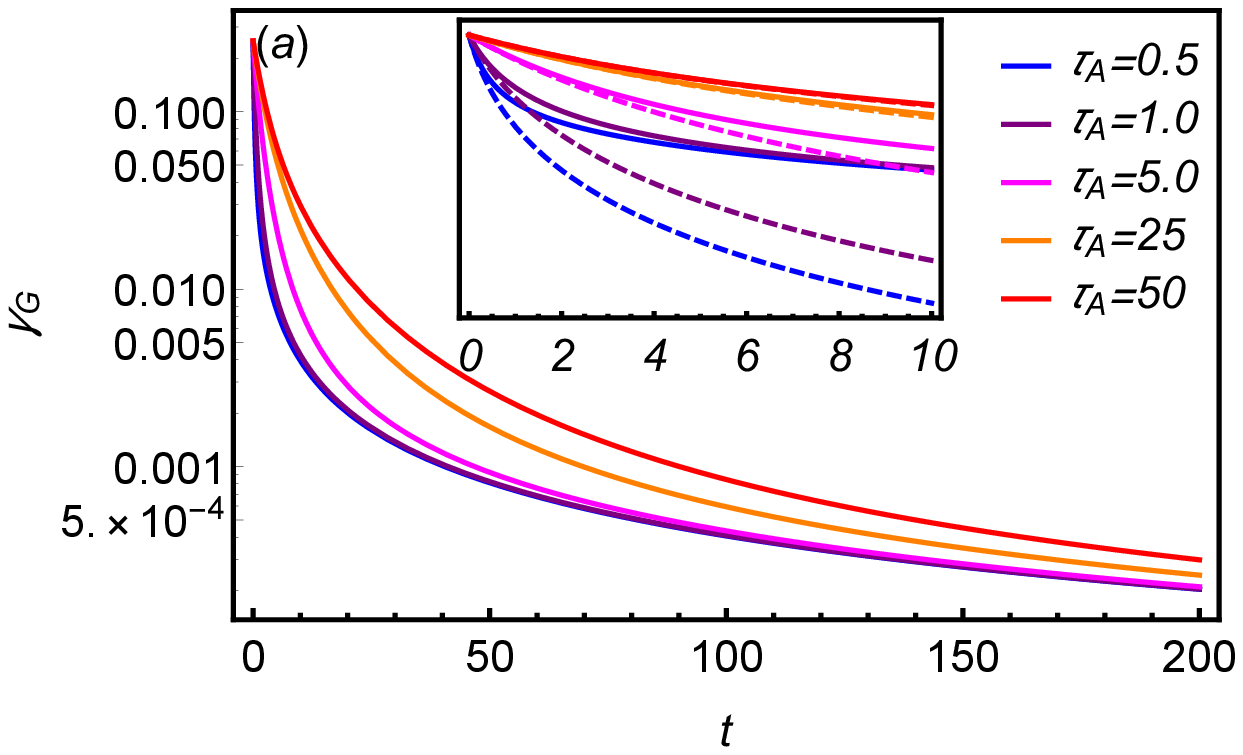}
\includegraphics[width=0.45\linewidth]{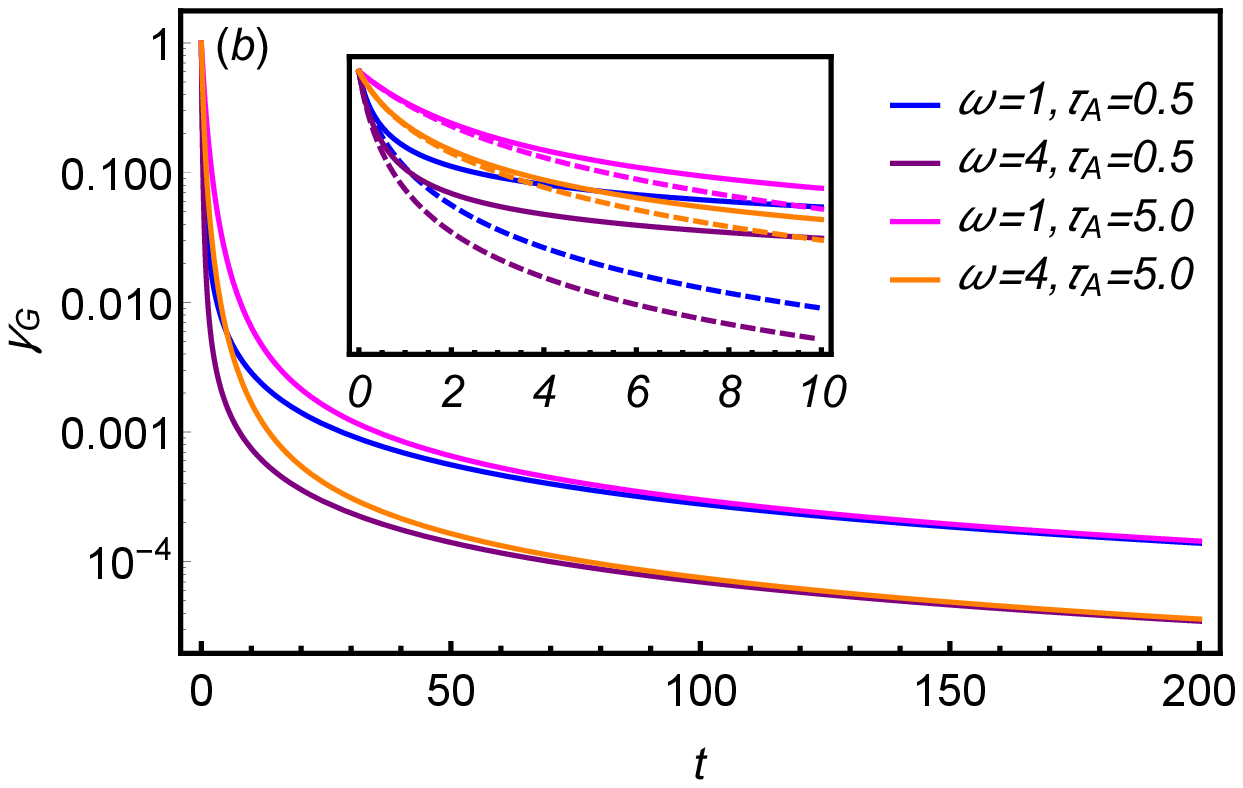}
\includegraphics[width=0.45\linewidth]{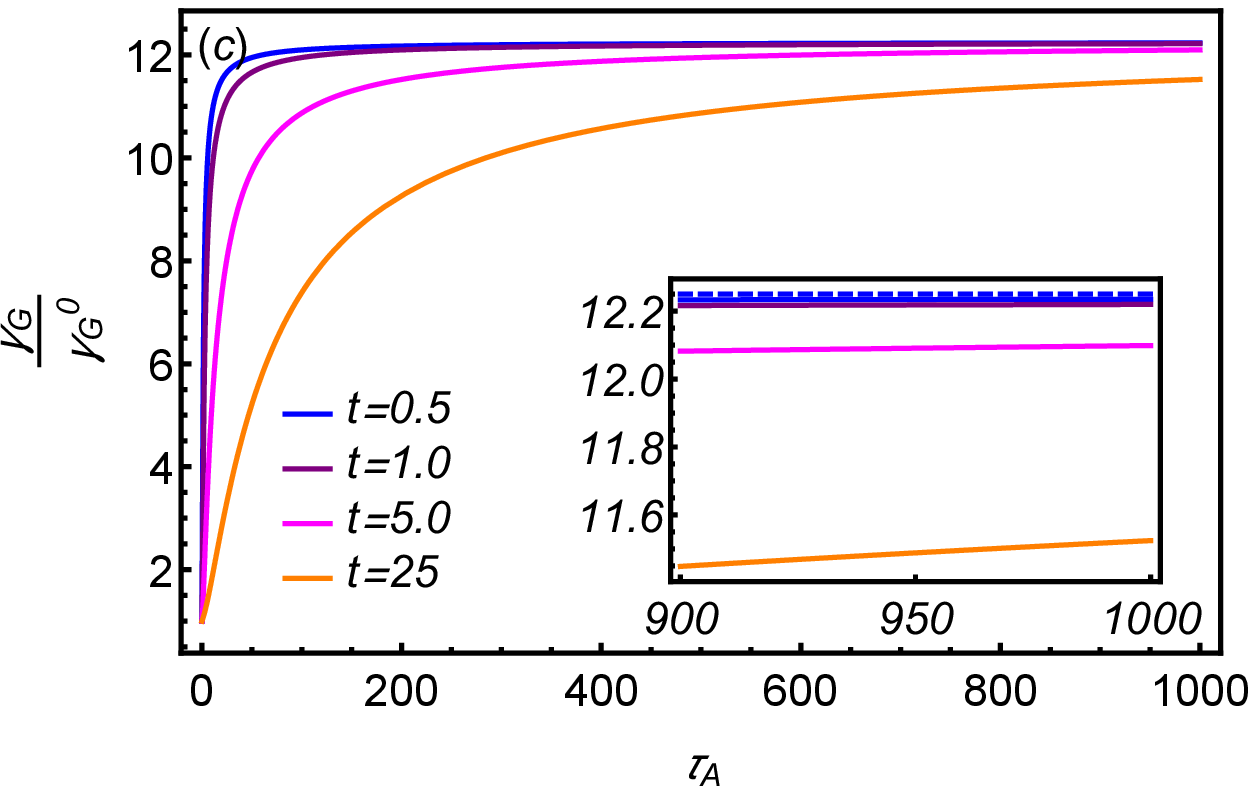}
\includegraphics[width=0.45\linewidth]{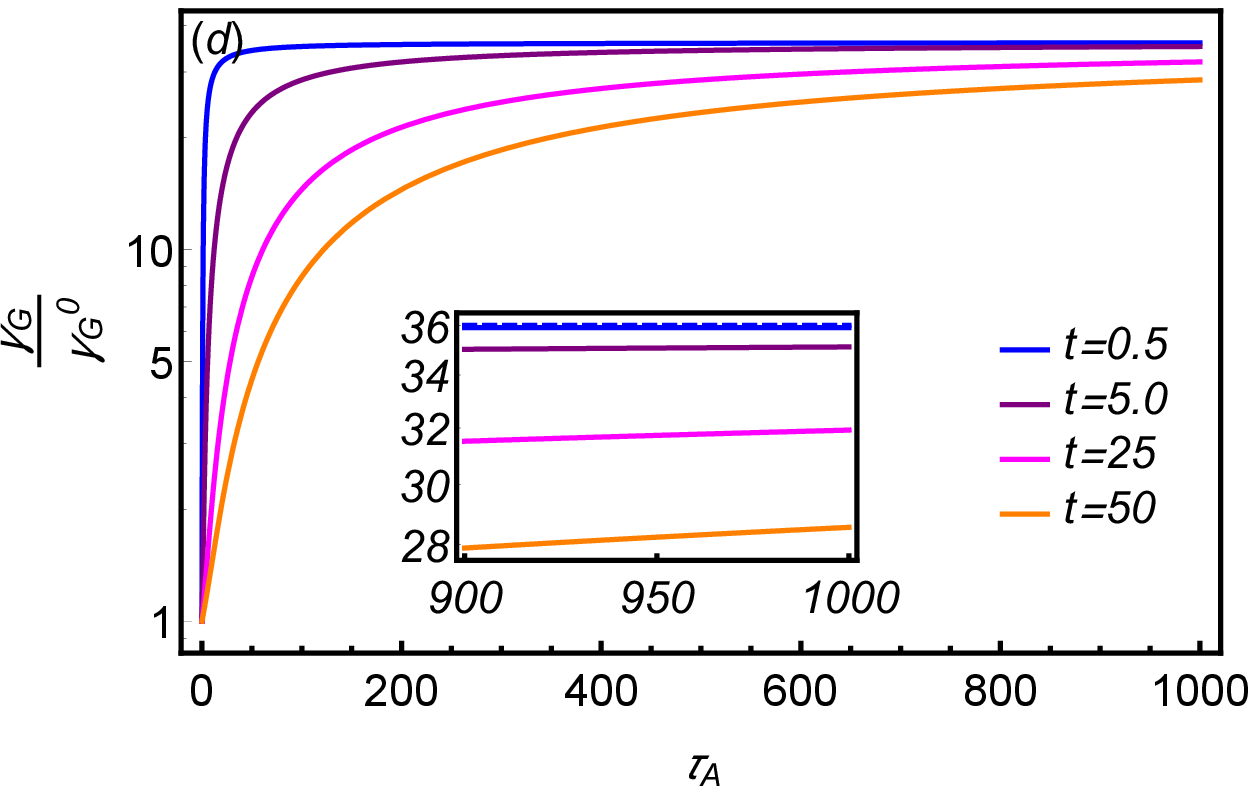}
\caption{The non-Gaussian parameter (NGP) is plotted (i) as a function of time $t$ for different values of $\tau_A$ in panels (a)-(b), and (ii)  the plots of NGP scaled by its value at $\tau_A=0$ [see Eqs. (\ref{ngp10}) and (\ref{ngp20})] are shown in panels (c)-(d) as a function of the correlation time of the active noise $\tau_A$ at different times.  In panels (a) and (c), the thermal diffusivity is modelled by the switching diffusion, and the curves correspond to Eq. (\ref{ngp1}).  The parameters used here are given by the set \{$D_1=5,\,D_2=15,\,D_A=25,\,r_{12}=r_{21}=1/2$\}. In panels (b) and (d), the diffusing diffusivity model is used to describe the fluctuating diffusivity, and the curves are obtained using Eq. (\ref{ngp2}). Here,   the values of other parameters are $D_{eq}=5,\,D_A=25.$ For panel (d), we take $\omega=1.0.$ In the insets of panels (a) and (b), the plots  are compared with the approximate results for $t\ll \tau_A$ (represented by the dashed lines) given by Eqs. (\ref{ngp_approxOD}) and (\ref{ngp_approxDD}), respectively. The plots for large-$\tau_A$ limit are enlarged in insets of panels (c) and (d) to show that the ratio $\gamma_G/\gamma^0_G$ converges to a constant value [see Eq. (\ref{approx_largetauOD})] which is represented by the dashed lines.   }
\label{ngp-pl}
 \end{figure}

\begin{figure}
 \centering
\includegraphics[width=0.5\linewidth]{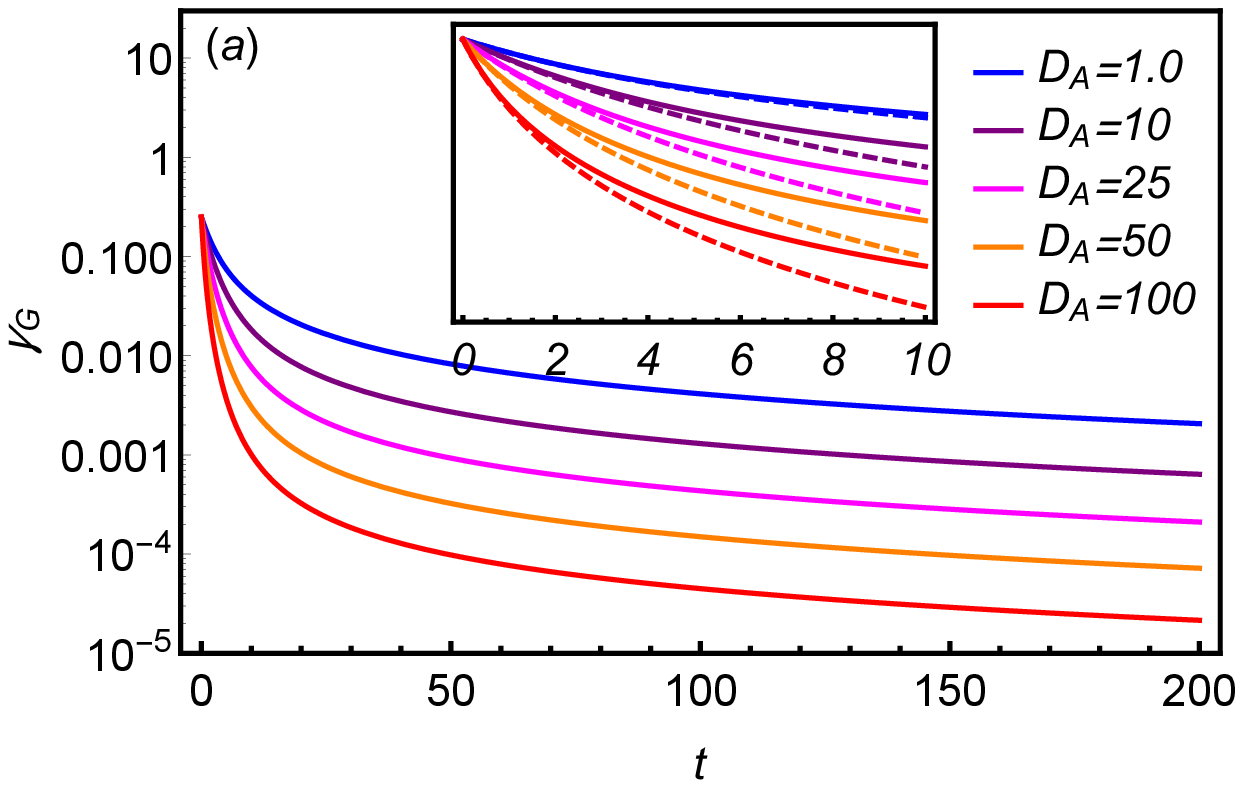}
\includegraphics[width=0.5\linewidth]{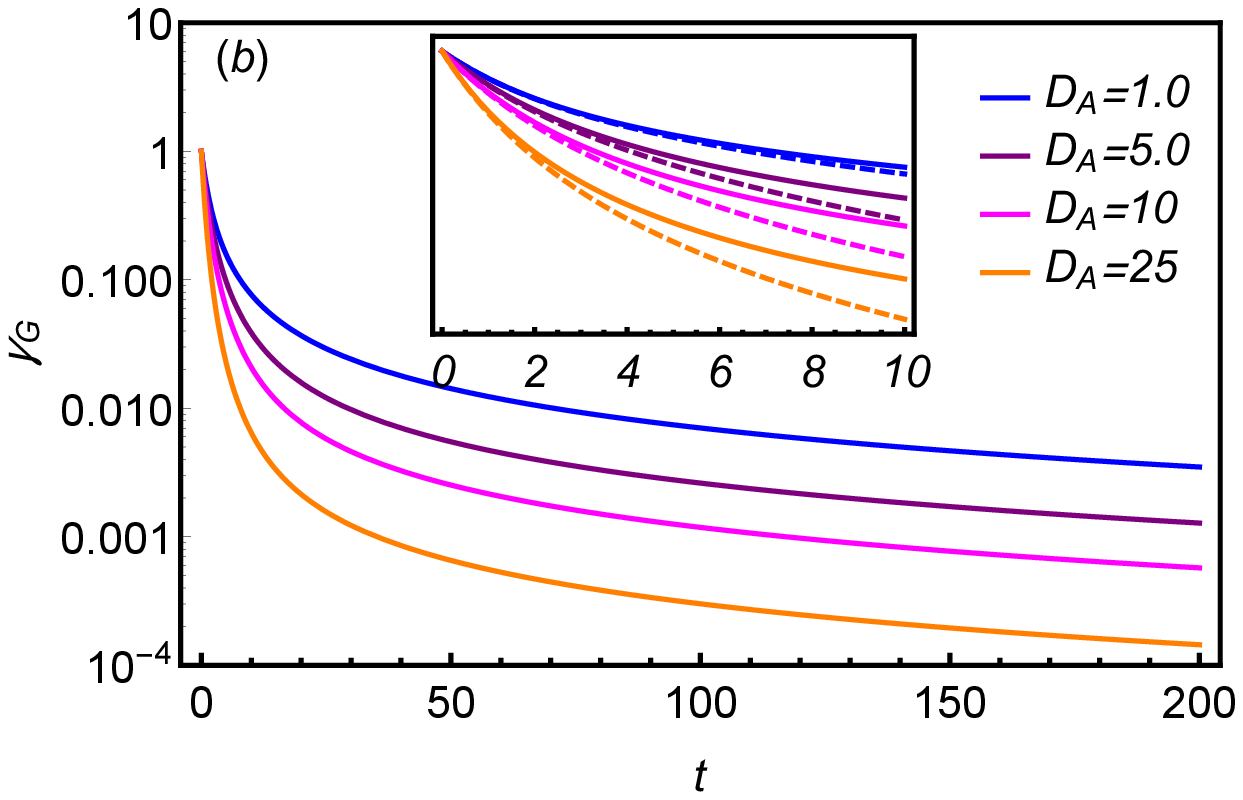}
\caption{Logarithmic pots of the non-Gaussian parameter (NGP) as a function of time $t$ for different values of $D_A$ in the case of (a)  switching diffusion and (b)  diffusing diffusivity model. The short-time behavior is shown in the inset and can be well described by the approximate results (represented by dashed lines) given by Eqs. (\ref{ngp_approxOD}) and (\ref{ngp_approxDD}), respectively. For both models, we have taken $\tau_A=5.$ The other parameters used in panel (a) are $D_1=5,\,D_2=15,\,r_{12}=r_{21}=\frac12,$ and in panel (b) these are $D_{eq}=5,\,\omega=1.$ }\label{ngp-DA}
 \end{figure}

\noindent For the switching diffusion, the NGP is given by (see the derivation in Appendix \ref{ngpcalc_OD})
\begin{align}
 & \gamma_G(t)= \frac{2 \left(D_2-D_1\right){}^2}{\left(r_{12}+r_{21}\right)^4}  \frac{ r_{12} r_{21} \left(\left(r_{12}+r_{21}\right) t-1+e^{-t \left(r_{12}+r_{21}\right)}\right) }{ \left( D_A \left(t-\tau _A\right)+D_{eq} t+e^{-\frac{t}{\tau _A}} D_A \tau _A\right)^2}.\label{ngp1}
\end{align}
 In the regime $t \ll \tau_A,$ it approximates to 
\begin{align}
   \gamma_G(t) \approx \frac{2 \left(D_2-D_1\right){}^2 r_{12} r_{21} \left(\left(r_{12}+r_{21}\right) t-1+e^{-t \left(r_{12}+r_{21}\right)}\right) }{\left(r_{12}+r_{21}\right)^4 \left(D_{eq} t+\frac12 D_A \frac{t^2}{\tau _A}\right)^2},\label{ngp_approxOD}
\end{align}
which implies $\gamma_G(t) \propto \tau_A^2$ at short times. This can be understood from Fig. \ref{ngp-pl} (a). In the extremely short-time limit $t\rightarrow 0,$ or equivalently for $1\gg r_{12} t,\,1\gg r_{21}t,\,t \ll \tau_A ,$ the NGP converges to a fixed value $\frac{\left(D_2-D_1\right){}^2 r_{12} r_{21}}{\left(r_{12}+r_{21}\right)^2 D_{eq}^2 },$ which is independent of $\tau_A$ and $D_A,$ as can be seen in Figs. \ref{ngp-pl} (a) and \ref{ngp-DA} (a). This indicates that the particle was initially at a nonequilibrium state solely due to  disorder. A quick check for the accuracy of our result is to take $D_2=D_1,$ and notice that the NGP vanishes to zero at short times, which correctly reproduces the result for  the case  without disorder.  As time passes, $\gamma_G(t)$ decreases monotonically, and it decays faster for small values of $\tau_A$ and large values of $D_A,$  as illustrated in in Figs. \ref{ngp-pl} (a) and \ref{ngp-DA} (a).  From Fig. \ref{ngp-pl} (c) one can also see that the NGP is a monotonically increasing function of $\tau_A,$ the slope  in the short  $\tau_A$ limit gradually decreases as $t$ increases, though its effect at  intermediate $\tau_A$ prevails only at a relevant timescale.  In the long-time limit, the NGP can be approximated to 
\begin{align}
\gamma_G(t) \approx  \frac{2 \left(D_2-D_1\right)^2 r_{12} r_{21} }{\left(r_{12}+r_{21}\right)^3 t \left( D_{eq}+D_A\right)^2},\label{ngp_approxOD1}
\end{align}
and it becomes zero at $t\rightarrow \infty,$ implying the Gaussian behavior, as anticipated.

\noindent For the diffusing diffusivity model, the NGP is given by (for detailed calculation see \ref{ngpcalc_DD})
\begin{align}
 \gamma_G(t) =\frac{D_{\text{eq}}^2 \left( 2 t \omega -1+e^{-2 t \omega }\right)}{2 \left(\omega t \left(D_A+D_{\text{eq}}\right)-D_A \omega \tau _A+\omega  D_A \tau _A e^{-\frac{t}{\tau _A}}\right)^2 }\label{ngp2}.
\end{align}
The NGP becomes unity and zero in the very short  and large time limits, respectively. In the limit $t\ll \tau_A,$ it can be approximated to 
\begin{align}
    \gamma_G(t) \approx \frac{D_{\text{eq}}^2 \left( 2 t \omega -1+e^{-2 t \omega }\right)}{2 \omega^2 \left( D_{\text{eq}} t+\frac12 D_A t^2/  \tau _A\right)^2 } .\label{ngp_approxDD}
\end{align}
 Like the previous case, the value of $\gamma_G(t)$ strongly depends on $\tau_A$ as well as $D_A,$ as can be seen in panels  (b) and (d) of Fig. \ref{ngp-pl} and in Fig. \ref{ngp-DA} (b). Also, similar dependence of $\omega$ on the NGP can be found, as shown in Fig. \ref{ngp-pl} (b). As the environment rearranges fast when $\omega$ is large, the motion is sampled over all realizations  at a shorter timescale and consequently, the particle achieves a Gaussian distribution, which is reflected in the plot of $\gamma_G(t)$ shown in Fig. \ref{ngp-pl} (b).

\begin{figure}
    \centering
    \includegraphics[width=0.5\linewidth]{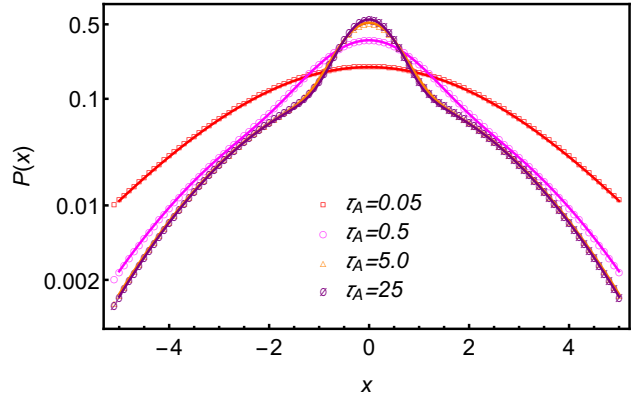}
\caption{Active particle with switching diffusion: spatial distribution computed numerically from  Eq. (\ref{P_2}) is plotted with symbols as a function of displacement $x$ for different values of $\tau_A$ at a short time $t=0.10.$ The values of other parameters are as follows: $D_1=1,\,D_2=14,\,r_{12}=1/2,\,r_{21}=1/2,\,D_A=25.$ The solid lines which correspond to the approximate result given in Eq. (\ref{analytic10}) are in good agreement with the numerical ones. }
    \label{pdfOD_tau}
\end{figure}

\begin{figure}
    \centering
    \includegraphics[width=0.5\linewidth]{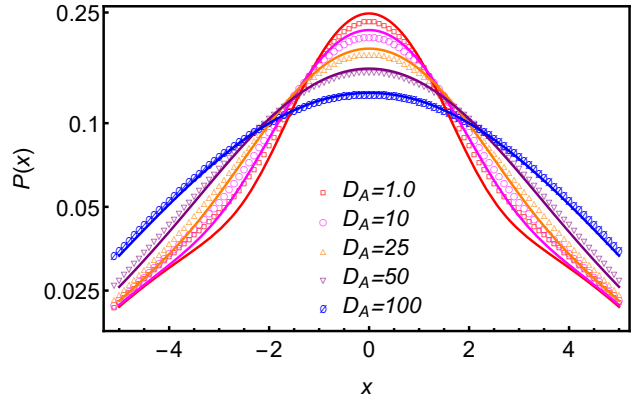}
\caption{Active particle with switching diffusion: PDF versus displacement plot for $t=0.50$ and $\tau_A=5.0.$  The values of other parameters ($D_1,\,D_2,\,r_{12},\,r_{21}$) are same as in Fig.   \ref{pdfOD_tau}, and the solid lines representing the approximate result [Eq. (\ref{analytic10})] are drawn for the purpose of comparison like Fig.   \ref{pdfOD_tau}. }\label{pdfOD_DA}
\end{figure}

\begin{figure}
 \centering
\includegraphics[width=0.5\linewidth]{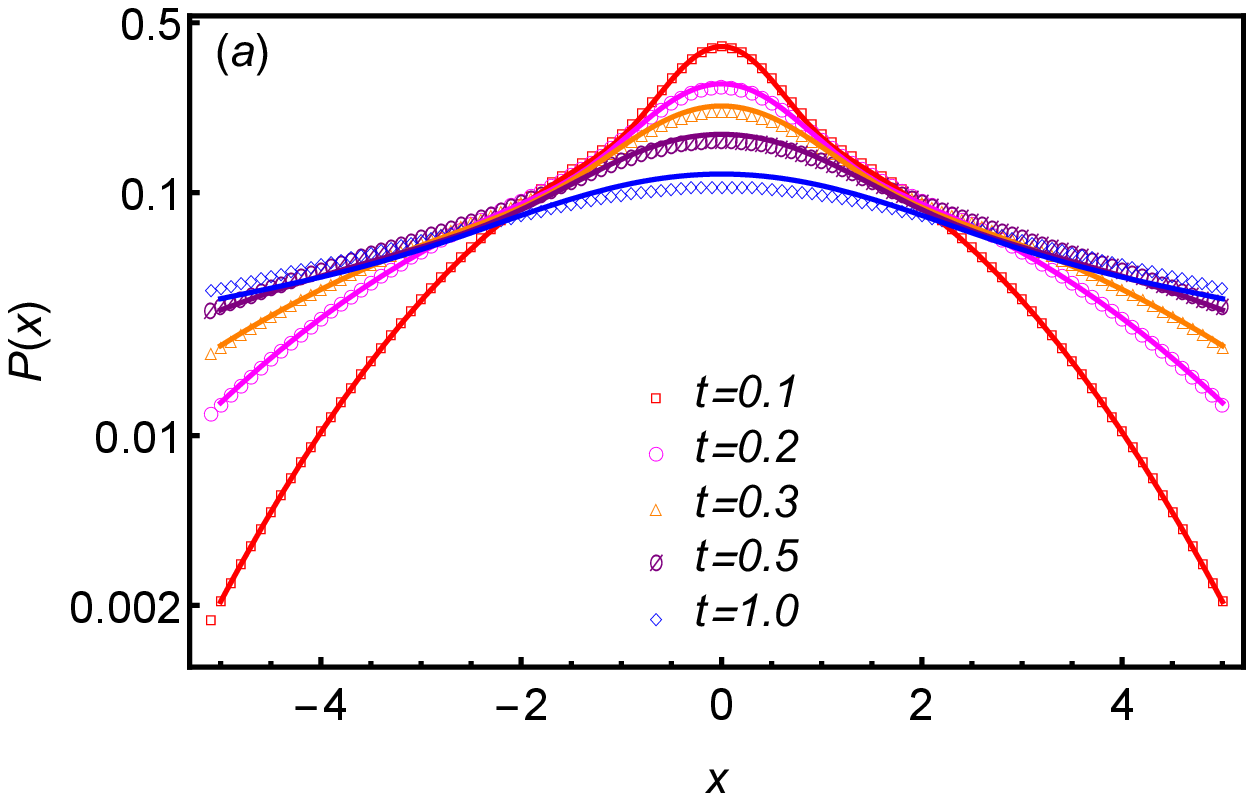}
\includegraphics[width=0.5\linewidth]{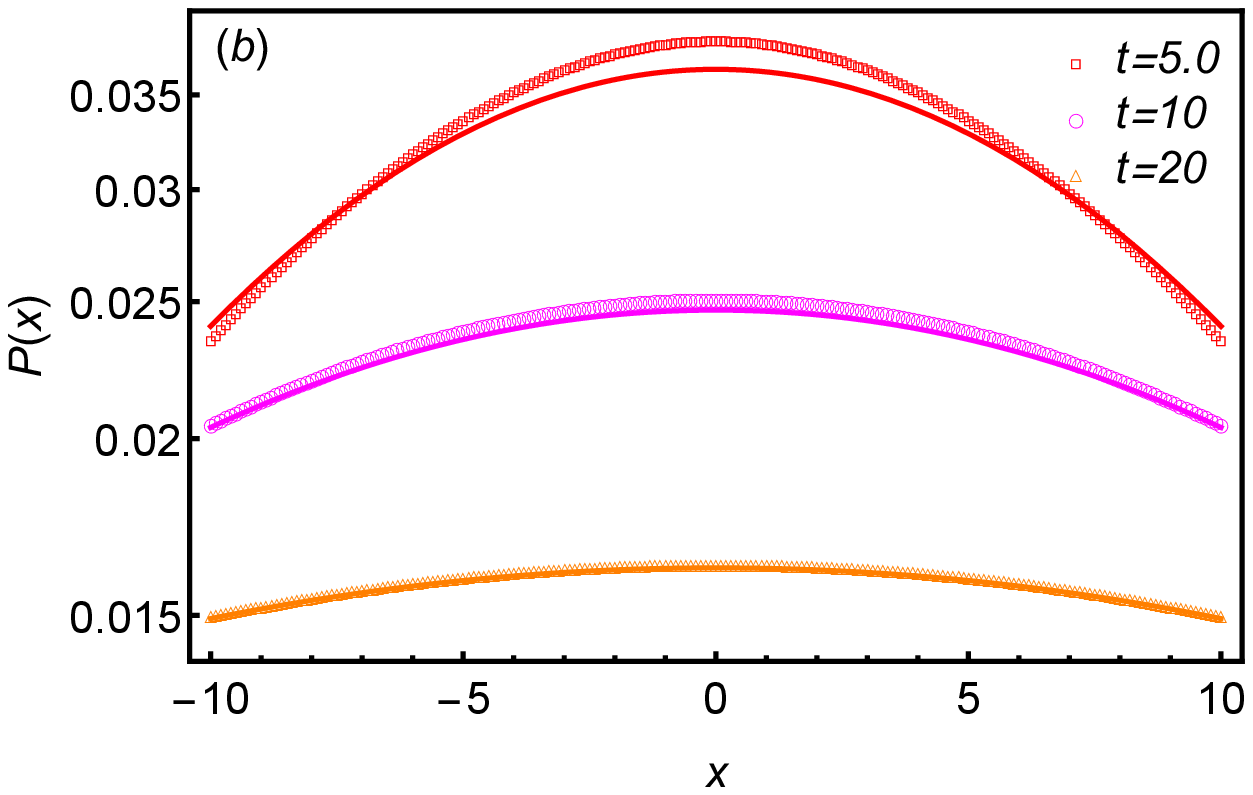}
\caption{Probability distribution function versus displacement in logarithmic scale at  different times for the case where the thermal diffusivity switches between two values $D_1$ (state 1) and $D_2$ (state 2) at Poisson rates (from state 2 to 1) $r_{21}$ and (vice versa) $r_{12}.$  The curves with symbols are obtained numerically using Eq. (\ref{P_2}) by taking the set of parameters as follows:  $\{D_1=1,\,D_2=14,\,r_{12}=3/4,\,r_{21}=1/4,\,\tau_A=50,\,D_A=25\}.$  In panel (a) the distribution is for small times and is approximately  similar to  Eq. (\ref{analytic10}) as depicted by solid lines.  In panel (b) the solid lines correspond to Eq. (\ref{analytic2}) and match well with the numerical results in the long-time limit.}
\label{pdfOD_times}
 \end{figure}

\noindent \paragraph{Switching diffusion:} Here we further investigate the dynamical properties by finding the complete PDF for the model of switching diffusion. By virtue of Eqs. (\ref{characterisctic-dicho}) and (\ref{ch_active}), the PDF given in Eq. (\ref{P_AOUP1}) can be written explicitly as 
\begin{align}
 &P(x,t|x_0=0,0)\nonumber\\&=\frac{1}{2\pi}\int_{-\infty}^{\infty}dp\, e^{-i p x} \frac{1}{2}\left[1-\frac{\phi_3}{\phi_2}\right]\,e^{-(\phi_1+\phi_2)t}e^{- D_A p^2 \left(t+\tau _A \left(e^{-\frac{t}{\tau _A}}-1\right)\right)}\nonumber\\
 &\quad+\frac{1}{2\pi}\int_{-\infty}^{\infty} dp\, e^{-i p x}\,\frac{1}{2}\left[1+\frac{\phi_3}{\phi_2}\right]\,e^{-(\phi_1-\phi_2)t}\,e^{- D_A p^2 \left(t+\tau _A \left(e^{-\frac{t}{\tau _A}}-1\right)\right)}\label{P_2}.
\end{align}
The integration in the above equation cannot be evaluated exactly. However, we can consider different limiting cases as  demonstrated in Appendix \ref{appn-OD}. At a timescale shorter  than the inverse of switching rates, the distribution is the weighted average of two Gaussian functions centered at $x_0,$ as given by (see Eq. (\ref{analytic11}))
\begin{align}
 P(x,t|x_0,0)=p_{1,s}\frac{e^{-\frac{(x-x_0)^2}{2\langle x^2(t)\rangle_1}}}{\sqrt{2\pi \langle x^2(t)\rangle_1}} +p_{2,s}\frac{e^{-\frac{(x-x_0)^2}{2\langle x^2(t)\rangle_2}}}{\sqrt{2\pi \langle x^2(t)\rangle_2}} \label{analytic10},
\end{align}
with $\langle x^2(t)\rangle_1=2\left[D_1\, t+D_A \left(t+\tau _A \left(e^{-\frac{t}{\tau _A}}-1\right)\right)\right]$, $\langle x^2(t)\rangle_2=2\left[D_2\, t+D_A \left(t+\tau _A \left(e^{-\frac{t}{\tau _A}}-1\right)\right)\right]$ and $p_{i,s}$ being the steady-state probabilities defined in Eq. (\ref{weight_st}).   So it shows a non-Gaussian behavior with a narrow central peak region which becomes pronounced at higher values of $\tau_A$ and lower values of $D_A,$  as can be seen in Figs. \ref{pdfOD_tau} and \ref{pdfOD_DA}.  These results are in consistent with the previous analysis on the non-Gaussian parameter, and it certainly requires a word of explanation. In our model, the dynamics solely driven by the active noise always shows a Gaussian behavior due to the Gaussian characteristics of the noise. However, with the introduction of dynamical disorder in the form of fluctuating thermal diffusivity, the system tends to deviate from the Gaussianity at short and intermediate timescales. Therefore, there is a competition between two opposing factors which determine the (non-)Gaussianity of the system. Naturally, prominent Gaussian behavior dictated by broader Gaussian central region prevails if the strength of the active noise given by the ratio $D_A/\tau_A$ becomes large. So using this argument all the previous results on the impact of $D_A$ and $\tau_A$ can be inferred. With the progress of time, the height of the peak decreases and the distribution flattens and eventually, it converges to a single Gaussian as  illustrated in Fig.  \ref{pdfOD_times}.   The Gaussian has the following form
\begin{align}
 P(x,t|x_0=0,0)=\frac{e^{-\frac{x^2}{2\langle x^2(t)\rangle_{eq}}}}{\sqrt{2\pi \langle x^2(t)\rangle_{eq}}},  \label{analytic2}
\end{align}
with $\langle x^2(t)\rangle_{eq}=2\left[D_{eq}\, t+D_A \left(t+\tau _A \left(e^{-\frac{t}{\tau _A}}-1\right)\right)\right].$ The calculations for this limiting case has been given in Appendix  \ref{appn-OD}. The above result is not very surprising as in the long-time limit the particle explores all possible realizations with different diffusivities, thereby displaying  Gaussian properties with an equilibrium diffusivity.  Unlike the  passive-particle case, here the width $\langle x^2(t)\rangle_{eq}$ is enhanced, and it is expressed in terms of  the equilibrium thermal diffusivity and active diffusivity. Note that here the pattern of convergence to a Gaussian is strikingly different from the one observed in a system with strong static disorder where the peak narrows down to a single point in the  long time  \cite{PhysRevLett.127.120601}. Our model is  related to the one in Ref. \cite{D1SM01507A} where a soft colloid diffusing in response to  the gradients in chemical potential and/or concentrations switches between two states with different masses due to active interactions. Unlike  Ref. \cite{D1SM01507A}, here we consider the motion of a soft active colloid, which means that the dynamics is not only driven by the chemical or/and concentration gradient, but it also moves due to self-propulsion mechanism fueled by ATP hydrolysis. However, it is interesting to note that both studies capture the non-Gaussian behavior at a shorter timescale, which further emphasizes the fact that the  non-Gaussianity is a very common feature associated with a disordered system.

\begin{figure}
 \centering
\includegraphics[width=0.5\linewidth]{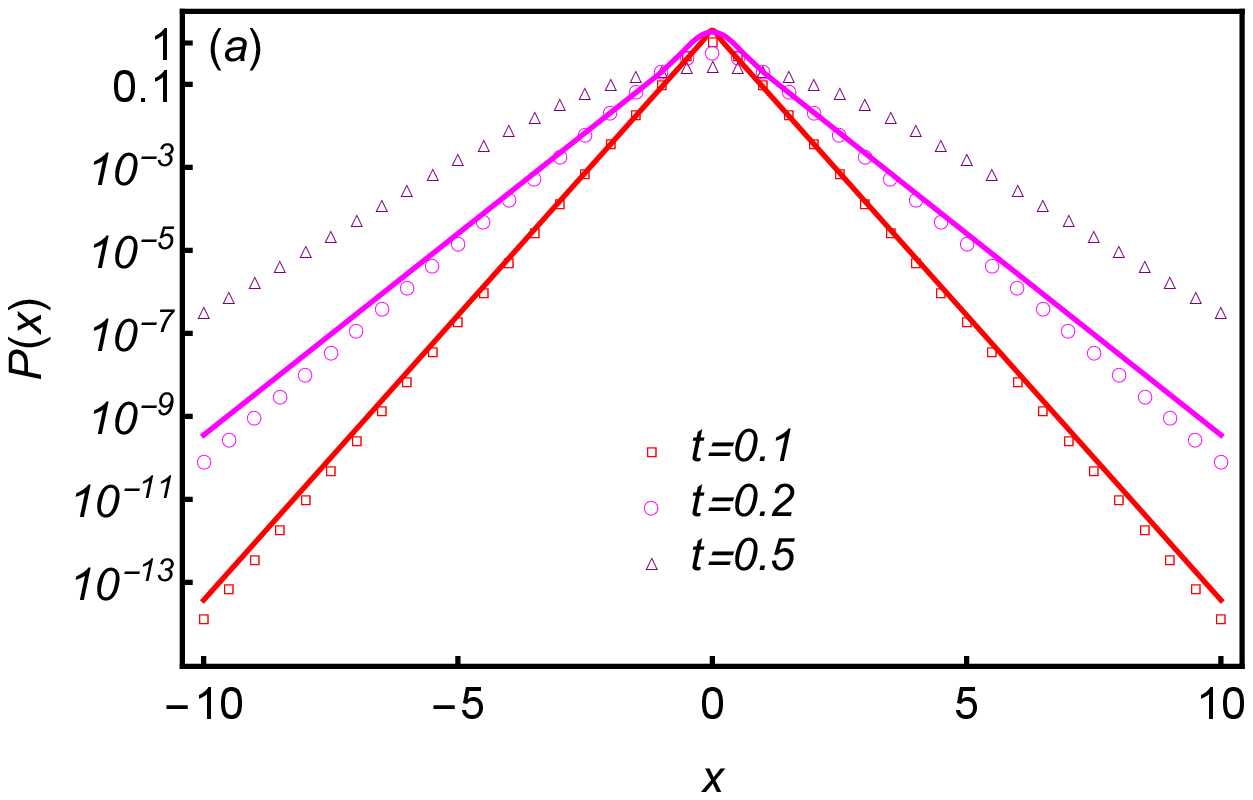}
\includegraphics[width=0.5\linewidth]{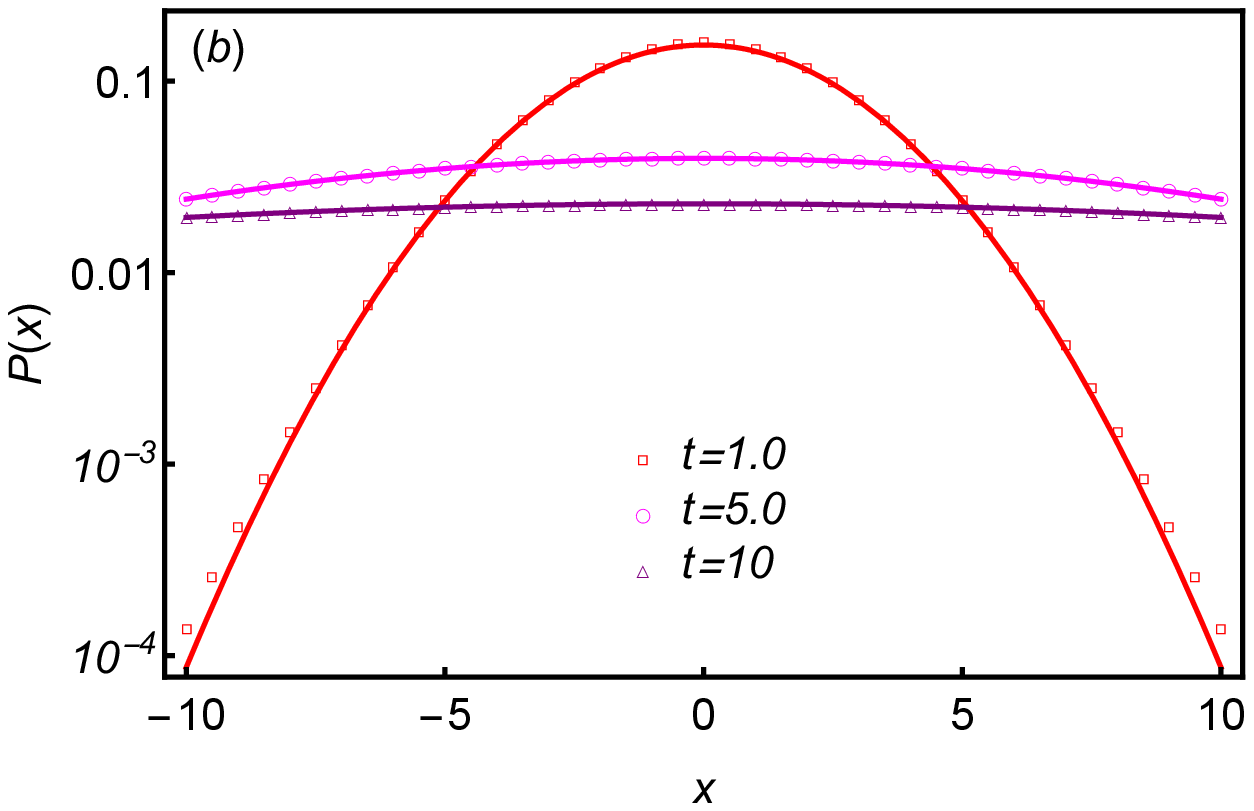}
\caption{Logarithmic plot of probability distribution function at  different times as a function of displacement  considering that the thermal diffusivity $D(t)$ is a stochastic variable following the OU process. The characteristic timescales for thermal diffusivity and active noise are $1/\omega=0.4,\,\tau_A=5.0,$ respectively. The numerical results obtained from  Eq. (\ref{P_AOUP_DD}) are plotted using symbols taking $D_A=25.$   In panel (a) the plots are for small times,  and the solid lines correspond to the analytical  result  given by Eq. (\ref{smallt_ana1}).  In panel (b) the solid lines correspond to Eq. (\ref{larget_ana1}).}
\label{msd-plOUP}
 \end{figure}

\begin{figure}
    \centering
    \includegraphics[width=0.5\linewidth]{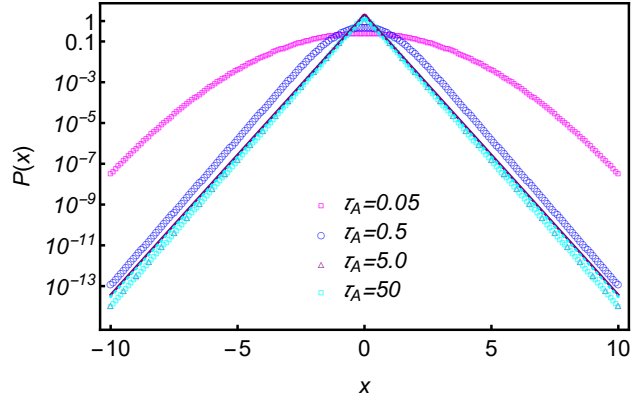}
\caption{Active particle in a heterogeneous medium characterized by the diffusing diffusivity model: logarithmic  plots of the spatial distribution numerically obtained from Eq. (\ref{P_AOUP_DD}) and marked by symbols are compared with the analytical solution [Eq. (\ref{smallt_ana1})] represented by the solid curves  for different values of $\tau_A$ at a short time $t=0.10.$ The values of other parameters are same as in Fig. \ref{msd-plOUP}. }
    \label{msd-plOUP1}
\end{figure}

\begin{figure}
    \centering
    \includegraphics[width=0.5\linewidth]{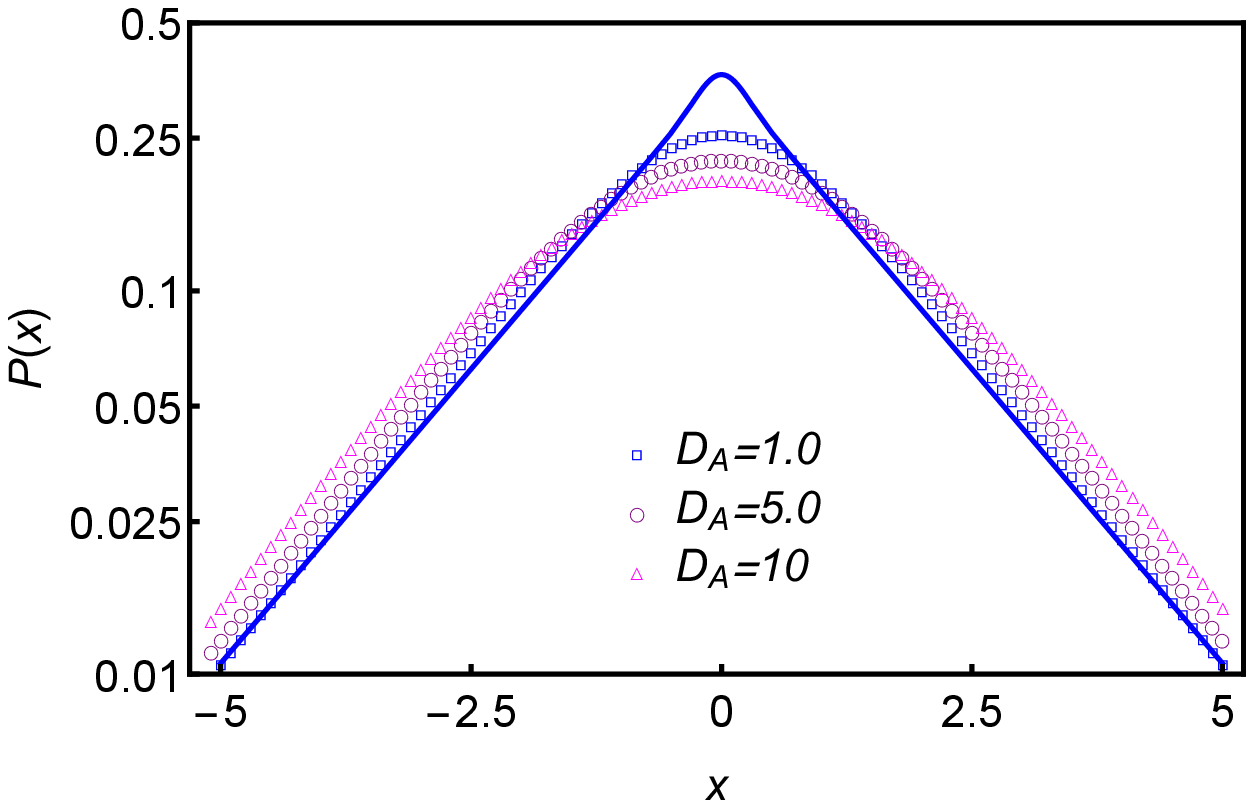}
\caption{Active particle in a heterogeneous medium characterized by the diffusing diffusivity model:  Like Fig. \ref{msd-plOUP1} the numerical plots of PDF  are compared with the approximate distribution (solid curve) given by Eq. (\ref{smallt_ana1}). The plots are for different values of $D_A$ at  time $t=0.40$ and $\tau_A=1.0.$ Other parameters are $D_{eq}=5$ and $\omega=1.0.$  }  \label{msd-plOUPD1}
\end{figure}

\noindent \paragraph{Diffusing diffusivity model: }
 From Eq. (\ref{P_AOUP1}), with aid of Eqs. (\ref{ch_active}) and (\ref{dangle}),  the spatial distribution can be expressed as
\begin{align}
&P(x,t|x_0=0,0)\nonumber\\
&=\frac{1}{2\pi}\int_{-\infty}^{\infty}dp\, e^{-i p x}\,\frac{4 \omega \beta\,e^{-(\beta-\omega) t}}{(\beta+\omega)^2-(\beta-\omega)^2 \,e^{-2\beta t}}\,e^{- D_A p^2 \left(t+\tau _A \left(e^{-\frac{t}{\tau _A}}-1\right)\right)}.\label{P_AOUP_DD}
 \end{align}
 The above integration is evaluated numerically and plotted in Figs. \ref{msd-plOUP} and \ref{msd-plOUP1}. Unlike the switching diffusion case, the distribution at short times is expressed by double exponential functions of the form 
 \begin{align}
 P&(x,t|x_0=0,0)\nonumber\\
 & \approx \frac{e^{\frac{D_A \left(t+\tau _A \left(e^{-\frac{t}{\tau _A}}-1\right)\right)}{D_{eq} t}}}{2\sqrt{ D_{eq} t }}\Bigl[\text{cosh}\left(\frac{x}{\sqrt{D_{eq} t}}\right)\nonumber\\
 & -\text{erf}\left(\frac{x}{\sqrt{\frac{1}{4 \left[D_A \left(t+\tau _A \left(e^{-\frac{t}{\tau _A}}-1\right)\right)\right] }}}\right)\text{sinh}\left(\frac{x}{\sqrt{D_{eq} t}}\right)\Bigr],\label{smallt_ana1}
 \end{align}
 which is also non-Gaussian in nature. See the derivation in Appendix \ref{appn-DD}. The non-Gaussian nature at short times can also be observed from Fig. \ref{msd-plOUP} (a). It is clear from  Fig.  \ref{msd-plOUP1} that the distribution  for large values of $\tau_A$  has a significantly long exponential tail, which indirectly suggests that the long persistence strongly enhances the effect of disorder on the particle's motion.  On the other hand, the nature of non-Gaussianity diminishes as the strength of active noise increases, as can  be comprehended from Fig. \ref{msd-plOUPD1}.  This is also complemented by the previous NGP analysis. Similar explanations as presented in the case of switching diffusion  can be given to understand such properties in the intermediate-time regime, namely, with increasing the strength of active noise the Gaussian properties prevail. As time passes, the height of the distribution gets reduced with flattening of the curve as shown in Fig. \ref{msd-plOUP} (b), and after a long time, the PDF becomes a Gaussian distribution given by Eq. (\ref{larget_ana1}). 

\section{Conclusion \label{conclusion}}
\noindent In this paper, we have investigated the dynamics of a single active Ornstein-Uhlenbeck particle (OUP)  subjected to the dynamical disorder. Invoking the idea of  fluctuating thermal diffusivity, we have shown that the distribution becomes non-Gaussian at intermediate times, which cannot be found in the usual OUP model, but the behavior is not very surprising as this is the signature of a disordered system. However, unlike the passive disordered case, the dynamics is not always Fickian; we have observed a ballistic behavior at a  timescale shorter than its persistence time, signifying the directed motion of the active particle. Also it has been found that the non-Gaussianity in the distribution is strongly influenced by the strength of the active noise suggesting that the longer correlation arising due to activity facilitates the effect of disorder on the dynamics.

\noindent The extensions of this work may be manyfold: In one direction, one can explore the effect of heterogeneity on several properties of a non-interacting active system which are at certain extent studied in the context of passive particle. In other direction, it can be extended to higher dimensions which is relevant while to be dealing with real biological systems. In addition, one may incorporate heterogeneity and interactions together to look into the many-body problem such as biopolymers and  active gels \cite{PhysRevE.92.032140,ahamad2020rouse}. 

\section{Conflicts of interest
}
There are no conflicts to declare.

\section{Acknowledgements}
\noindent  R. C. acknowledges SERB for funding (Project No. MTR/2020/000230
under MATRICS scheme) and IRCC-IIT Bombay (Project
No. RD/0518-IRCCAW0-001) for funding.
K.G. acknowledges IIT Bombay for support through the institute postdoctoral fellowship.

\appendix
\section{Computation of moments and non-Gaussian parameter}
\noindent The second moment of the distribution as given by Eq. (\ref{msd}) can be computed using Eqs. (\ref{white_corr}) and (\ref{color_corr}), and it reads
\begin{align}
&\langle (x(t)-x_0)^2\rangle\nonumber\\&=\int_{0}^{t}\int_{0}^{t}dt_1 dt_2\,\langle\eta_T(t_1)\eta_T(t_2)\rangle+\int_{0}^{t}\int_{0}^{t}dt_1 dt_2\,\langle v_A(t_1)v_A(t_2)\rangle \nonumber\\
&=2\int_{0}^{t}\int_{0}^{t}dt_1 dt_2\,\langle D_T(t_1)\rangle \delta(t_1-t_2)+\frac{D_A}{\tau_A} \int_{0}^{t}\int_{0}^{t}dt_1 dt_2\,e^{-\frac{|t_1-t_2|}{\tau_A}}\nonumber\\
&=2\langle D_T(t)\rangle\int_{0}^{t} dt_1 +2D_A \int_{0}^{t} dt_1\,\left(1-e^{-\frac{t_1}{\tau_A}}\right)\nonumber\\
&=2\,D_{eq}\, t+2\,D_A \left(t+\tau_A \left(e^{-\frac{t}{\tau _A}}-1\right)\right)\label{msd11}.
\end{align}

\noindent The fourth moment given in Eq. (\ref{fourthmonent}) can be expressed using Eq. (\ref{xt}), as follows:
\begin{align}
&\langle (x(t)-x_0)^4\rangle\nonumber\\&=\int_{0}^{t}\int_{0}^{t}\int_{0}^{t}\int_{0}^{t}dt_1 dt_2 dt_3 dt_4\,\langle\eta_T(t_1)\eta_T(t_2) \eta_T(t_3)\eta_T(t_4)\rangle \nonumber\\
&\;+2\int_{0}^{t}\int_{0}^{t}\int_{0}^{t}\int_{0}^{t}dt_1 dt_2dt_3 dt_4\,\langle \eta_T(t_1)\eta_T(t_2)\rangle \langle v_A(t_3) v_A(t_4)\rangle\nonumber\\
&\;+\int_{0}^{t}\int_{0}^{t}\int_{0}^{t}\int_{0}^{t}dt_1 dt_2dt_3 dt_4\,\langle v_A(t_1)v_A(t_2) v_A(t_3) v_A(t_4)\rangle. \label{fourthmonent1}
\end{align}
Now, with the aid of Eq. (\ref{fourthmonent}) and the autocorrelation function of $v_A(t)$ as given in Eq. (\ref{color_corr}), the  last term of the RHS in the  above equation can be computed, and it reads 
\begin{align}
   &\int_{0}^{t}\int_{0}^{t}\int_{0}^{t}\int_{0}^{t}dt_1 dt_2dt_3 dt_4\,\langle v_A(t_1)v_A(t_2) v_A(t_3) v_A(t_4)\rangle\nonumber\\
   &=\int_{0}^{t}\int_{0}^{t}dt_1 dt_2 \langle v_A(t_1)v_A(t_2)\rangle   \int_{0}^{t}\int_{0}^{t}dt_3 dt_4 \langle v_A(t_3) v_A(t_4)\rangle\nonumber\\&+\int_{0}^{t}\int_{0}^{t}dt_1 dt_3 \langle v_A(t_1)v_A(t_3)\rangle \int_{0}^{t}\int_{0}^{t}dt_2 dt_4 \langle v_A(t_2) v_A(t_4)\rangle\nonumber\\ 
   &+\int_{0}^{t}\int_{0}^{t}dt_1 dt_4 \langle v_A(t_1)v_A(t_4)\rangle \int_{0}^{t}\int_{0}^{t}dt_2 dt_3 \langle v_A(t_2) v_A(t_3)\rangle\nonumber\\
   &=3\left[\frac{D_A}{\tau_A} \int_{0}^{t}\int_{0}^{t}dt_1 dt_2 e^{-\frac{|t_1-t_2|}{\tau_A}}\,\times\,\frac{D_A}{\tau_A} \int_{0}^{t}\int_{0}^{t}dt_3 dt_4 e^{-\frac{|t_3-t_4|}{\tau_A}}\right]\nonumber\\
   &=12 \left[D_A \left(\tau_A  \left(e^{-\frac{t}{\tau_A }}-1\right)+t\right)\right]^2.\label{term3}
\end{align}
Using the properties of white noise $\eta_T(t),$ the first term of the RHS in Eq. (\ref{fourthmonent1}) can be written as 
\begin{align}
&\int_{0}^{t}\int_{0}^{t}\int_{0}^{t}\int_{0}^{t}dt_1 dt_2 dt_3 dt_4\,\langle\eta_T(t_1)\eta_T(t_2) \eta_T(t_3)\eta_T(t_4)\rangle\nonumber\\&=\int_{0}^{t}\int_{0}^{t} \int_{0}^{t}\int_{0}^{t} dt_1 dt_2 dt_3 dt_4 \langle\langle \eta_T(t_1)\eta_T(t_2) \rangle  \langle \eta_T(t_3) \eta_T(t_4)\rangle\rangle_{D_T}\nonumber\\
  & +\int_{0}^{t}\int_{0}^{t}  \int_{0}^{t}\int_{0}^{t}dt_1 dt_3 dt_2 dt_4 \langle\langle\eta_T(t_1)\eta_T(t_3)\rangle \langle \eta_T(t_2)\eta_T(t_4)\rangle\rangle_{D_T}\nonumber\\ 
   &+\int_{0}^{t}\int_{0}^{t} \int_{0}^{t}\int_{0}^{t} dt_1 dt_4 dt_2 dt_3 \langle\langle \eta_T(t_1)\eta_T(t_4)\rangle  \langle \eta_T(t_2) \eta_T(t_3)\rangle\rangle_{D_T}\nonumber\\
   &=24 \int_{0}^{t} dt_1\int_{0}^{t_1} dt_2 \langle D_T(t_1)D_T(t_2)\rangle.\label{lasrterm}
\end{align}
The middle term of the RHS in Eq. (\ref{fourthmonent1}) is given by 
\begin{align}
&\int_{0}^{t}\int_{0}^{t}\int_{0}^{t}\int_{0}^{t}dt_1 dt_2dt_3 dt_4\,\langle \eta_T(t_1)\eta_T(t_2)\rangle \langle v_A(t_3) v_A(t_4)\rangle\nonumber\\
&=12t \int_{0}^{t}dt_1\int_{0}^{t_1}dt_2\,\langle D(t_1)\rangle\langle v_A(t_1) v_A(t_2)\rangle.\label{midterm}
\end{align}
For the two models of $D_T(t),$ we  solve Eqs.  (\ref{lasrterm}) and (\ref{midterm}). The calculations are  shown below.
\subsection{Switching diffusion \label{ngpcalc_OD}}
\noindent To perform the integration in Eqs.  (\ref{lasrterm}) and (\ref{midterm}), it is required to know the average  and autocorrelation function of $D(t)$ which    are given by \cite{balakrishnan1993simple} 
 $\langle  D(t)\rangle=\frac{r_{12}D_2+r_{21}D_1}{r_{12}+r_{21}},$ and $\langle  D(t_1) D(t_2)\rangle=\langle  D(t_1)\rangle^2+\frac{r_{12}r_{21}}{(r_{12}+r_{21})^2}(D_2-D_1)^2\,e^{-(r_{12}+r_{21})|t_2-t_1|}.$
 Using these, one can compute Eq. 
\ref{lasrterm} as 
\begin{align}
  &\int_{0}^{t}\int_{0}^{t}\int_{0}^{t}\int_{0}^{t}dt_1 dt_2 dt_3 dt_4\,\langle\eta_T(t_1)\eta_T(t_2) \eta_T(t_3)\eta_T(t_4)\rangle \nonumber\\
  &=24 \int_{0}^{t} dt_1\int_{0}^{t_1} dt_2\Bigl[\left(\frac{r_{12}D_2+r_{21}D_1}{r_{12}+r_{21}}\right)^2\nonumber\\
  &\quad\quad\quad\quad\quad\quad+\frac{r_{12}r_{21}}{(r_{12}+r_{21})^2}(D_2-D_1)^2\,e^{-(r_{12}+r_{21})(t_1-t_2)}\Bigr]\nonumber\\
  &=\frac{12 }{\left(r_{12}+r_{21}\right){}^3} \Bigl[\left(r_{12}+r_{21}\right) t^2 \left(D_2 r_{12}+D_1 r_{21}\right){}^2\nonumber\\
  &\quad\quad\quad\quad\quad\quad\quad +2 \left(D_1-D_2\right){}^2 r_{12} r_{21}\left(\frac{e^{\left(-r_{12}-r_{21}\right) t}-1}{r_{12}+r_{21}}+t\right)\Bigr].\label{terms1}
\end{align}
Similarly, one has 
\begin{align}
&2\int_{0}^{t}\int_{0}^{t}\int_{0}^{t}\int_{0}^{t}dt_1 dt_2dt_3 dt_4\,\langle \eta_T(t_1)\eta_T(t_2)\rangle \langle v_A(t_3) v_A(t_4)\rangle\nonumber\\
&=  \frac{24\, D_A t \left(D_2 r_{12}+D_1 r_{21}\right) e^{-\frac{t}{\tau_A }} \left(e^{t/\tau_A } (t-\tau_A )+\tau_A \right)}{r_{12}+r_{21}}. \label{terms2} 
\end{align}
  By virtue of Eqs. (\ref{term3}), (\ref{terms1}),  (\ref{terms2}) and (\ref{msd}), the  NGP defined by 
 $\gamma_G=\frac{ \langle x^4(t)\rangle}{3\langle x^2(t)\rangle^2}-1$ 
  can be calculated, and it yields
\begin{align}
 & \gamma_G(t)= \frac{2 \left(D_2-D_1\right){}^2}{\left(r_{12}+r_{21}\right)^2} \times\nonumber\\
 & \frac{ r_{12} r_{21} \left(\left(r_{12}+r_{21}\right) t-1+e^{-t \left(r_{12}+r_{21}\right)}\right) }{ \left(\left(r_{12}+r_{21}\right) D_A \left(t-\tau _A\right)+(D_2 r_{12} +D_1 r_{21}) t+e^{-\frac{t}{\tau _A}}\left(r_{12}+r_{21}\right) D_A \tau _A\right)^2}.\label{ngp11}
\end{align}
For $\tau_A=0,$ the NGP denoted here as $\gamma^0_G(t)$ can be expressed as 
\begin{align}
 \gamma^0_G(t)= \frac{2 \left(D_2-D_1\right){}^2 r_{12} r_{21} \left(\left(r_{12}+r_{21}\right) t-1+e^{-t \left(r_{12}+r_{21}\right)}\right) }{\left(r_{12}+r_{21}\right)^4 t^2 \left( D_A +D_{eq}\right)^2}.  \label{ngp10} 
\end{align}
 For $\tau_A\rightarrow \infty$ and finite values of $t,$ the ratio $\gamma_G(t)/\gamma^0_G(t)$ takes a constant value, $viz.,$ 
\begin{align}
   \frac{\gamma_G(t)}{\gamma^0_G(t)} \approx \left(1+\frac{D_A}{D_{eq}}\right)^2.\label{approx_largetauOD}
\end{align}

\subsection{Diffusing diffusivity model \label{ngpcalc_DD}}
\noindent Using the dynamical equation of $\xi(t)$ given in Sec. \ref{DD_model}, one can  write 
\begin{align}
&\langle\left(\xi(t_1)-\xi(0) e^{-\omega t_1}\right)\left(\xi(t_2)-\xi(0) e^{-\omega t_2}\right)\rangle\nonumber\\
&=\langle\xi(t_1)\xi(t_2)\rangle\nonumber\\
&\;-\langle\xi(t_1)\xi(0)\rangle e^{-\omega t_2}-\langle\xi(t_2)\xi(0)\rangle e^{-\omega t_1}+\langle\xi(0)\xi(0)\rangle e^{-\omega t_1-\omega t_2}\nonumber\\
&=\int_{0}^{t_1}\int_{0}^{t_2}dt'_1dt'_2\, e^{-\omega(t_1-t'_1)}e^{-\omega(t_2-t'_2)}\langle\eta_A(t'_1)\eta_A(t'_2)\rangle,
\end{align}
which implies $\langle\xi(t_1)\xi(t_2)\rangle=\langle\xi^2(0)\rangle e^{-\omega|t_1-t_2|}.$ With these, one can get 
\begin{align}
&\sum_{i,j=1}^{2}\langle\xi_i^2(t_1)\xi_j^2(t_2)\rangle=\sum_{i,j=1,\,i\neq j}^{2}\langle\xi_i^2(t_1)\rangle\langle\xi_j^2(t_2)\rangle+ \sum_{i=1}^{2}  \langle\xi^2_i(t_1)\xi^2_i(t_2)\rangle\nonumber\\
&\quad\quad\quad\quad\quad\quad\quad=2\langle\xi^2(0)\rangle^2+ \sum_{i=1}^{2} \left( \langle\xi_i^2(t_1)\rangle^2+2\langle\xi_i(t_1)\xi_i(t_2)\rangle^2 \right)\nonumber\\
&\quad\quad\quad\quad\quad\quad\quad=4\langle\xi^2(0)\rangle^2+4 \langle\xi^2(0)\rangle^2  e^{-2\omega|t_1-t_2|}.\label{xit}
\end{align}
Here, $\langle D(t) \rangle = \sum_{n=1}^{2} \xi_n^2(t),$ and $\langle \xi_i^2(t)\rangle=\langle \xi_i^2(0)\rangle=\frac12 D_{eq}.$
Using Eq. (\ref{xit}), Eq. (\ref{lasrterm})  can be calculated as follows:
\begin{align}
&\int_{0}^{t}\int_{0}^{t}\int_{0}^{t}\int_{0}^{t}dt_1 dt_2 dt_3 dt_4\,\langle\eta_T(t_1)\eta_T(t_2) \eta_T(t_3)\eta_T(t_4)\rangle\nonumber\\
   &=24 \int_{0}^{t}dt_1\int_{0}^{t_1} dt_2 \langle D_T(t_1)D_T(t_2)\rangle\nonumber\\
   &=24  \int_{0}^{t}dt_1\int_{0}^{t_1} dt_2 \sum_{i,j=1}^{2}\langle\xi_i^2(t_1)\xi_j^2(t_2)\rangle\nonumber\\
   &=24 \int_{0}^{t}dt_1\int_{0}^{t_1} dt_2\, \frac14\langle \xi^2(t_1)\rangle^2+24 \int_{0}^{t}dt_1\int_{0}^{t_1} dt_2\,\frac14 \langle\xi^2(0)\rangle^2  e^{-2\omega|t_1-t_2|}\nonumber\\
   &=\frac{6\, D_{\text{eq}}^2 \left[2 t \omega  (t \omega +1)+e^{-2 t \omega }-1\right]}{\omega ^2}.\label{terms11}
\end{align}
Similarly, from Eq. (\ref{midterm}) one can obtain 
\begin{align}
&2\int_{0}^{t}\int_{0}^{t}\int_{0}^{t}\int_{0}^{t}dt_1 dt_2dt_3 dt_4\,\langle \eta_T(t_1)\eta_T(t_2)\rangle \langle v_A(t_3) v_A(t_4)\rangle\nonumber\\
&=24t \int_{0}^{t}dt_1\int_{0}^{t_1}dt_2\,\langle D(t_1)\rangle\langle v_A(t_1) v_A(t_2)\rangle\nonumber\\
&=24 D_{eq} D_A\,t\left(\tau _A \left(e^{-\frac{t}{\tau _A}}-1\right)+t\right).\label{terms21}
\end{align}
 Using Eqs. (\ref{term3}), (\ref{terms11}),  (\ref{terms21}) and (\ref{msd}), the  NGP can be calculated, and it reads 
\begin{align}
 \gamma_G(t) =\frac{ \langle x^4(t)\rangle}{3\langle x^2(t)\rangle^2}-1=\frac{D_{\text{eq}}^2 \left( 2 t \omega -1+e^{-2 t \omega }\right)}{2 \left(\omega t \left(D_A+D_{\text{eq}}\right)-D_A \omega \tau _A+\omega  D_A \tau _A e^{-\frac{t}{\tau _A}}\right)^2 }\label{ngp21}.
\end{align}
From the above equation one can find the  NGP at $\tau_A=0,$ which reads
\begin{align}
 \gamma^0_G(t) =\frac{ \langle x^4(t)\rangle}{3\langle x^2(t)\rangle^2}-1=\frac{D_{\text{eq}}^2 \left( 2 t \omega -1+e^{-2 t \omega }\right)}{2 \omega^2 t^2 \left(D_A+D_{\text{eq}}\right)^2 }\label{ngp20}.
\end{align}
Like the previous case, the ratio $\gamma_G(t)/\gamma^0_G(t)$ for $\tau_A\rightarrow \infty$ and finite values of $t$ is well approximated to Eq.  (\ref{approx_largetauOD}).

\section{Limiting values of PDF}
\noindent Here we show the calculation of approximate PDFs at different limits. Two models of the fluctuating thermal diffusivity are considered for the analysis.
\subsection{Switching diffusion \label{appn-OD}}
\noindent Here one can assign two diffusive timescales  $D_2 p^2$ and $D_1 p^2,$ to compare them with the switching rates $r_{12}$ and $r_{21}.$ For  $\frac{D_2}{r_{21}}p^2\gg 1$ and $\frac{D_1}{r_{12}}p^2\gg 1,$ one gets 
$\phi_2 \approx \frac{1}{2}\left[(D_2-D_1)p^2 +(r_{21}-r_{12})\right]+\frac{r_{12}r_{21}}{(D_2-D_1)p^2 +(r_{21}-r_{12})}.$  Clearly, it  corresponds to the liming case at shorter times.  With the above approximation,  Eq. (\ref{P_2})  modifies to 
\begin{align}
& P(x,t|x_0,0)\nonumber\\
&=\frac{1}{2\pi}\int_{-\infty}^{\infty}dp\, e^{-i p (x-x_0)} \frac{r_{12}}{r_{12}+r_{21}}\,e^{-D_2\,p^2 t}e^{- D_A p^2 \left(t+\tau _A \left(e^{-\frac{t}{\tau _A}}-1\right)\right)}\nonumber\\    
 &\;+\frac{1}{2\pi}\int_{-\infty}^{\infty} dp\, e^{-i p (x-x_0)} \frac{r_{21}}{r_{12}+r_{21}}\,e^{-D_1\,p^2t}e^{- D_A p^2 \left(t+\tau _A \left(e^{-\frac{t}{\tau _A}}-1\right)\right)}\nonumber\\
 &=\frac{r_{12}}{r_{12}+r_{21}}\,\frac{e^{-\frac{(x-x_0)^2}{4\left[D_2\, t+D_A \left(t+\tau _A \left(e^{-\frac{t}{\tau _A}}-1\right)\right)\right]}}}{\sqrt{4\pi\left[D_2\, t+D_A \left(t+\tau _A \left(e^{-\frac{t}{\tau _A}}-1\right)\right)\right]}}\nonumber\\&
 +\frac{r_{21}}{r_{12}+r_{21}}\,\frac{e^{-\frac{(x-x_0)^2}{4\left[D_1\, t+D_A \left(t+\tau _A \left(e^{-\frac{t}{\tau _A}}-1\right)\right)\right]}}}{\sqrt{4\pi\left[D_1\, t+D_A \left(t+\tau _A \left(e^{-\frac{t}{\tau _A}}-1\right)\right)\right]}}\label{analytic11}.
\end{align}
 Notice that, within the short-time  regime one can consider two limiting cases,  $r_{12}\gg r_{21}$ and $r_{12}\ll r_{21}.$ For $r_{12}\gg r_{21},$  the first term on the RHS  in Eq. (\ref{analytic11}) dominates and the second term can be ignored, and therefore, the distribution becomes almost Gaussian with a width containing only  $D_2$ and $D_A.$ In the other limiting case, the behavior is mostly dictated by the second term.   
In the long-time limit,   $i.e.,$  for $\frac{D_2}{r_{21}}p^2\ll 1$ and $\frac{D_1}{r_{12}}p^2\ll 1,$ the following approximations can be done, 
$\phi_2\approx \frac12 (r_{12}+r_{21})-\frac{p^2}{2} \frac{r_{12}-r_{21}}{r_{12}+r_{21}}(D_2-D_1),$ and
$\phi_1-\phi_2\approx p^2\left(\frac{r_{12}}{r_{12}+r_{21}}D_2+\frac{r_{21}}{r_{12}+r_{21}}D_1\right) \approx p^2\,D_{eq}.$
Therefore, Eq. (\ref{P_2}) approximates to 
\begin{align}
P(x,t|x_0,0) = \frac{e^{-\frac{(x-x_0)^2}{4\left[D_{eq}\, t+D_A \left(t+\tau _A \left(e^{-\frac{t}{\tau _A}}-1\right)\right)\right]}}}{\sqrt{4\pi\left[D_{eq}\, t+D_A \left(t+\tau _A \left(e^{-\frac{t}{\tau _A}}-1\right)\right)\right]}}. \label{analytic21}
\end{align}
\subsection{Diffusing diffusivity model \label{appn-DD}}
\noindent  In the short-time limit, $i.e.,$ for  $\beta t\ll 1,$ and for small and intermediate displacements ($i.e.$, $D_{eq}p^2/\omega\gg 1$), the above can be approximated as 
 \begin{align}
 P&(x,t|x_0=0,0) \approx \frac{1}{2\pi}\int_{-\infty}^{\infty}dp \frac{e^{-i p x}}{1+D_{eq}p^2 t}\,e^{- D_A p^2 \left(t+\tau _A \left(e^{-\frac{t}{\tau _A}}-1\right)\right)}\nonumber\\
 &\approx \frac{1}{2\sqrt{ D_{eq} t }}\sqrt{\frac{1}{4\pi \left[D_A \left(t+\tau _A \left(e^{-\frac{t}{\tau _A}}-1\right)\right)\right] }}\nonumber\\
 &\int_{-\infty}^{\infty}dx'\,e^{-\frac{|x-x'|}{\sqrt{ D_{eq} t }}}\,e^{-\frac{x'^2}{4 \left[D_A \left(t+\tau _A \left(e^{-\frac{t}{\tau _A}}-1\right)\right)\right] }}\nonumber\\
 & \approx \frac{e^{\frac{D_A \left(t+\tau _A \left(e^{-\frac{t}{\tau _A}}-1\right)\right)}{D_{eq} t}}}{2\sqrt{ D_{eq} t }}\text{cosh}\left(\frac{x}{\sqrt{D_{eq} t}}\right)\nonumber\\
 & -\frac{e^{\frac{D_A \left(t+\tau _A \left(e^{-\frac{t}{\tau _A}}-1\right)\right)}{D_{eq} t}}}{2\sqrt{ D_{eq} t }}\text{erf}\left(\frac{x}{\sqrt{\frac{1}{4 \left[D_A \left(t+\tau _A \left(e^{-\frac{t}{\tau _A}}-1\right)\right)\right] }}}\right)\text{sinh}\left(\frac{x}{\sqrt{D_{eq} t}}\right).\label{smallt_ana111}
 \end{align}
 In the long $t$ limit, $i.e.,$ for  $\beta t\gg 1,$ the PDF [Eq. (\ref{P_AOUP_DD})] can be approximated as 
 \begin{align}
 & P(x,t |x_0=0,0)  \approx \frac{1}{2\pi}\int_{-\infty}^{\infty} dp e^{-i p x}\,e^{-D_{eq}p^2 t}e^{- D_A p^2 \left(t+\tau _A \left(e^{-\frac{t}{\tau _A}}-1\right)\right)}\nonumber\\
 &\approx \sqrt{\frac{1}{4\pi \left[D_{eq} t+D_A \left(t+\tau _A \left(e^{-\frac{t}{\tau _A}}-1\right)\right)\right] }}e^{-\frac{x^2}{4 \left[D_{eq} t+D_A \left(t+\tau _A \left(e^{-\frac{t}{\tau _A}}-1\right)\right)\right] }} \label{larget_ana1}.
 \end{align}




\providecommand*{\mcitethebibliography}{\thebibliography}
\csname @ifundefined\endcsname{endmcitethebibliography}
{\let\endmcitethebibliography\endthebibliography}{}

\end{document}